\documentclass[aps,prx,twocolumn,notitlepage,eqsecnum,
superscriptaddress,nofootinbib]{revtex4-2}

\usepackage[centertags]{amsmath} \usepackage{amssymb} \usepackage{latexsym}
\usepackage{enumerate} \usepackage{graphicx} \usepackage{mathrsfs}
\usepackage[colorlinks]{hyperref} \usepackage{stmaryrd} \usepackage{import}
\usepackage{tensor} \usepackage[usenames,dvipsnames]{xcolor} \usepackage{bm}
\usepackage{multirow} \usepackage{breakurl} \usepackage{float}
\usepackage{verbatim} \usepackage{mathrsfs}
\usepackage{booktabs}
\usepackage{orcidlink}
\usepackage[normalem]{ulem} 
\usepackage[capitalise]{cleveref}
\crefname{section}{Sec.}{Sec.}
\Crefname{section}{Section}{Sections}
\newcommand{\ccite}[1]{Ref.~\cite{#1}}
\newcommand{\ccites}[1]{Refs.~\cite{#1}}

\allowdisplaybreaks[1]

\definecolor{CiteColor}{rgb}{0,0.5,0} \hypersetup{citecolor=CiteColor}
\definecolor{RefColor}{rgb}{0.55,0,0} \hypersetup{linkcolor=RefColor}
\definecolor{darkgreen}{rgb}{0.2,0.7,0.2}
\definecolor{red}{rgb}{1,0,0}

\begin{document}

\title{Machine learning and high accuracy: Neural preconditioners for computational gravity}

\newcommand{\geneseo}{\affiliation{Department of Physics and Astronomy, State University of New York at Geneseo, New York 14454, USA}}
\newcommand{\ucd}{\affiliation{School of Mathematics and Statistics, University College Dublin, Belfield, Dublin 4, Ireland}}
\newcommand{\uzh}{\affiliation{Department of Astrophysics, University of Zurich, Winterthurerstrasse 190, 8057 Zurich, Switzerland}}

\author{Thomas Osburn\,\orcidlink{0000-0003-2747-3994}} \email{tosburn3@gmail.com} \geneseo \ucd
\author{Nils Vu\,\orcidlink{0000-0002-5767-3949}} \email{nils.vu@uzh.ch} \uzh

\date{\today}

\begin{abstract}

Machine learning (ML) developments have empowered valuable calculations in many areas of theoretical physics. Certain topics involving moderate-to-low accuracy requirements ($\sim1\%$) have been particularly fortunate because many ML techniques, such as neural networks, introduce a noise floor that would have been difficult to penetrate with stricter tolerances. For this reason, topics involving high accuracy requirements are sometimes hesitant to pursue ML. One prominent example is gravitational wave source modeling, which typically relies on traditional numerical methods that are able to accommodate the high sensitivities of upcoming detectors such as the Laser Interferometer Space Antenna (LISA). However, recent progress in applied mathematics has identified hybrid methods where fast-but-noisy ML mechanisms are inserted into otherwise traditional numerical schemes to accelerate convergence in a way that ultimately meets or exceeds the accuracy of standard methods.
We develop an optimized hybrid ML method to tackle a class of challenging elliptic partial differential equations (PDEs); specifically, we demonstrate ML implementations for predicting gravitational waves from asymmetric-mass black hole binaries and for modeling elastic waves. These require solving sparse linear algebra systems with Helmholtz-type features that are especially difficult for traditional iterative methods. Our new approach is to train a graph neural network (GNN) as a significantly faster preconditioner using a novel loss function based on the true residual of the iterative algorithm. 
For the most difficult regions of parameter space, our neural preconditioner decreases iteration counts by a factor of 2-15$\times$ compared to traditional preconditioners. 
Because the GNN operates directly on the sparse matrix and needs only modest physical context, it generalizes beyond its training domain and can be adapted to new problems across physics by retraining rather than manual redesign.
Hybrid ML methods such as this offer a new and compelling avenue to handle high accuracy and speed requirements in gravitational wave modeling and beyond.

\end{abstract}

\maketitle

\section{Introduction}
\label{sec:intro}

\begin{figure*}[t]
\includegraphics[trim={0 3 0 3},clip,width=\textwidth]{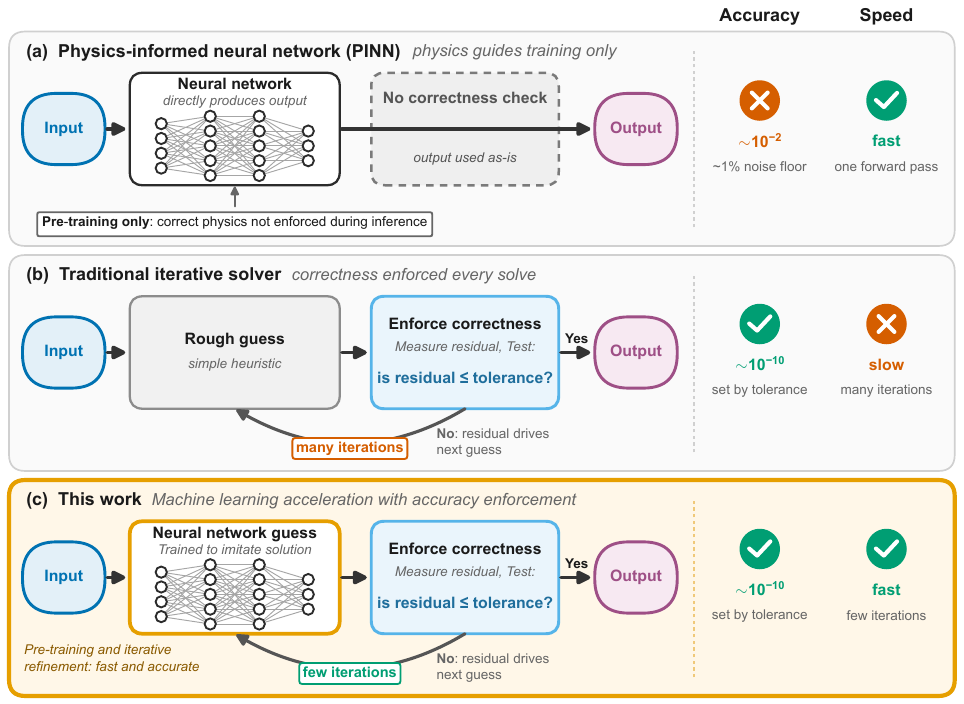}
\caption{It can be tricky to balance speed and accuracy when considering ML methods. (a) Physics-informed neural networks are fast, but introduce a relatively high noise floor. (b) Traditional iterative solvers assess accuracy each cycle, but often require a very large number of iterations to converge. (c) This work seeks the best of both worlds by implementing a hybrid ML method where fewer total iterations are required via training neural networks to optimize the amount of improvement per iteration.}
\label{fig:idea} 
\end{figure*}

Machine learning (ML) and artificial intelligence (AI) have empowered valuable calculations across physics~\cite{Carleo_2019,Karniadakis_2021}, from neural representations of quantum many-body states~\cite{Carleo_2017} and real-time control of fusion plasmas~\cite{Degrave_2022} to optimized weather predictions~\cite{Lam2023GraphCast,Kochkov2024NeuralGCM} and rapid approximations of partial differential equation (PDE) solutions~\cite{Raissi_2019,Li_2021}. Gravitational physics, the primary focus of this work, is no exception~\cite{Yoo_2023,Freitas_2025,cranganore2026}. However, many of these successful ML applications are enabled by physical contexts that tolerate lower accuracies (${\sim}1\%$). Popular ML tools, such as neural networks, typically plateau at a modest noise floor (see \cref{fig:idea}) and learn high-frequency features slowly, a phenomenon known as spectral bias~\cite{Rahaman_2019}. Physics-informed neural network (PINN) predictions in black hole perturbation theory and general relativity saturate near this noise level~\cite{Luna_2023,carretero2026}, with the most accurate direct ML applications in gravity reaching relative errors of ${\sim}10^{-4}$~\cite{Yoo_2026}. Furthermore, reported ML advantages for PDEs can be biased due to weak baselines~\cite{McGreivy_2024}. Consequently, fields with stringent accuracy requirements, such as gravitational wave source modeling for the Laser Interferometer Space Antenna (LISA)~\cite{LISA:2024hlh,LISAWaveformWG:2023arg}, can be hesitant to adopt ML (data analysis with ML is an exception~\cite{Rebei_2019,Haegel_2020,Dax_2021,Dax_2023,benedetto2023ai,pereira2024,Cuoco_2025}).

Although theoretical physics predictions can involve virtually any mathematics, this work focuses on an especially ubiquitous category: sparse linear systems. Common examples where sparse matrices arise include when elliptic PDEs are discretized, whenever stiff equations are evolved implicitly, and within every Newton iteration for non-linear problems~\cite{Knoll_2004}. Krylov methods such as GMRES~\cite{Saad_1986} are the workhorses for large sparse systems, but their convergence is largely determined by preconditioner performance, which must exploit problem-specific features~\cite{Benzi_2002}. Although certain comparatively simpler scenarios, such as electrostatic PDE calculations, have solved many of their preconditioning problems, several high-impact areas have struggled to reliably identify successful preconditioners. Magnetically confined fusion plasmas span time scales from fast Alfv\'en waves to slow resistive transport, and their implicit time stepping produces stiff, ill-conditioned systems where preconditioning remains a central challenge~\cite{Knoll_2004,Jardin_2012}. Optical inverse problems, such as nanophotonic inverse design, require repeated solutions of indefinite frequency-domain Maxwell systems for which Krylov methods converge slowly~\cite{Molesky_2018,Shin_2013}; other similar problems such as seismic full-waveform inversion face related issues~\cite{Virieux_2009} because preconditioning these Helmholtz-type problems remains a major open problem in numerical analysis~\cite{ErnstGander2012,Gander_2019,Erlangga2006}. In lattice QCD, inverting the Dirac operator dominates the cost and becomes severely ill-conditioned near physical quark masses, requiring years of specialized multigrid development~\cite{Babich_2010,Frommer_2014,Clark_2010}. Similar difficulties arise in radiation transport~\cite{Adams_2002} and numerical relativity initial data~\cite{Vu_2022}.

The problem of identifying and deploying high-performing preconditioners is ideal for ML because preconditioners are inherently imperfect; minor noise associated with ML-based preconditioning should be negligible because overall accuracy is guaranteed by the underlying Krylov framework, see \cref{fig:idea}. Training of ML preconditioners to adapt to their specific context is a powerful accelerator for iterative solvers; relevant techniques include learned multigrid~\cite{Luz_2020}, learned incomplete
factorizations~\cite{Li_2023,hausner2024,hausner2024GMRES,trifonov2025learninglinearalgebragraph}, graph neural preconditioners~\cite{Chen_2025}, neural operators within Krylov
and relaxation methods~\cite{pmlr-v235-rudikov24a,Zhang_2024,eshaghi2026nowsneuraloperatorwarm},
and Helmholtz-specific approaches~\cite{Azulay_2023}; see Ref.~\cite{meng2026overviewmachinelearningenhancediterative} for an overview. Note that lattice QCD has already begun using neural preconditioners~\cite{Lehner_2023,Lehner_2024,Cali_2023}, but most areas of physics involving sparse linear systems have not yet benefited from this powerful tool. Furthermore, many applied mathematics demonstrations target symmetric positive definite model problems rather than the indefinite, coupled systems typical of frontier physics. This work introduces a new and improved loss function and training strategy with applicability to multiple sparse problems; our method enables substantial iterative acceleration via graph neural network (GNN) preconditioners that learn an optimized matrix factorization~\cite{hausner2024,hausner2024GMRES}. 

Our primary application is asymmetric-mass compact binaries, where a black hole or neutron star inspirals towards a larger black hole. These extreme and intermediate mass-ratio systems produce vital LISA signals with ${\sim}10^2$-$10^6$ cycles~\cite{LISA:2024hlh,EMRI} and are modeled via black hole perturbation theory and gravitational self-force~\cite{Poisson:2011nh,Barack:2018yvs,Pound_2021}. Accurate parameter estimation requires non-linear self-force calculations through second-order in the mass-ratio~\cite{Hinderer_2008,Burke_2024}, and second-order results have so far been largely restricted to quasicircular orbits around non-spinning (Schwarzschild) black holes~\cite{Pound_2020,Warburton_2021,Wardell_2023}. Extending non-linear order self-force calculations to generic orbits around spinning (Kerr) black holes is an important open challenge. Although first-order self-force calculations for generic Kerr orbits have been successful~\cite{vandeMeent_2018,Nasipak_2019,Nasipak_2026} thanks largely to separability of the Teukolsky equation~\cite{Teukolsky_1973}, second-order involves problematic mode-coupling~\cite{Miller_2016,Spiers_2024} and slow polar convergence that are both exacerbated when using full separation of variables to calculate second-order Kerr perturbations. 

A promising approach for second-order black hole perturbations, and the primary focus of this work, is the frequency-domain $m$-mode method~\cite{Osburn_2022,Macedo_2024,Vu_2026,osburn2026schwarzschildperturbationslorenzgauge}, which was recently generalized to Kerr spacetime~\cite{nishimura2026,leather2026b,cownden2026,casals2026}. Separating only the $t$ and $\varphi$ dependence yields coupled 2D elliptic PDEs describing the $m$-modes of the spacetime metric perturbation. However, these PDEs are approximately of the Helmholtz type: terms proportional to $\omega^2$ introduce spatial modes that oscillate resonantly in a way that harms iterative convergence~\cite{ErnstGander2012,ChowSaad1997}. Attempts to avoid iterative convergence issues by using a direct matrix solver encounter prohibitive time and memory requirements (see Sec.~\ref{sec:results} and Ref.~\cite{Osburn_2022}). Second-order calculations for generic Kerr orbits will require PDE solves for vast numbers of frequencies, $m$-modes, sources, and spins, which implies acceleration of iterative convergence will be vital to enable comprehensive LISA waveform template coverage.

In this work we show how neural preconditioners can provide this acceleration without sacrificing accuracy. Interpreting each sparse matrix as the weighted adjacency matrix of a graph, we train GNNs to produce corrected incomplete LU (ILU) factorizations~\cite{hausner2024,hausner2024GMRES} that have been enhanced to significantly accelerate GMRES convergence. In contrast, prior training+loss strategies produced preconditioners even worse than basic ILU for our systems. Our new loss function directly measures GMRES residual reduction, and we backpropagate by differentiating through GMRES restart cycles. For the most challenging high-frequency black hole perturbation problems, the resulting preconditioner converges to relative residuals of $10^{-10}$ with 2-15$\times$ fewer GMRES iterations than basic ILU and 2-4$\times$ less wall time than the next best method (often the direct method, which is only competitive due to the modest resolutions studied here). This strategy eliminates the need for a memory-intensive direct solver, repairs near-zero pivots that can cause basic ILU to fail entirely, and generalizes beyond its training domain in frequency, $m$, resolution, and source features.

To demonstrate cross-disciplinary reach and interest, we also apply our approach to elastodynamics by considering time-harmonic elastic waves driven near the surface of a non-homogeneous solid, which generalizes Lamb's problem~\cite{Lamb_1904} and Gibson's soil model~\cite{Gibson_1967}. This problem is valuable for seismology~\cite{Virieux_2009,li20152d}, with further applications in materials science and engineering, and it inherits the difficulties of Helmholtz-type problems~\cite{Yovel_2024}. In addition, elasticity problems already have some overlap with gravitational wave science~\cite{Vu:2021mhw}. Even standard multigrid smoothers fail for shear-wave components in elastodynamics~\cite{Rizzuti_2016}. Li et~al.~\cite{li20152d} contrast the difficult elasticity elliptic problem with the more attainable acoustic elliptic problem: 
``An efficient preconditioner based on a complex shifted Laplacian operator has been successfully implemented for the acoustic wave equation. However, extensions to the elastic case have not been performed, to the best of our knowledge.'' 
After only minor customization, our neural preconditioner again outperforms traditional iterative methods. Because our GNN operates directly on the sparse matrix with only modest physical context, it can be adapted to new problems by retraining rather than by manually redesigning preconditioners. By this logic, neural
preconditioners such as the GNNs investigated here are likely to become increasingly powerful tools for quantitative problems involving sparse linear systems.

The remainder of this paper is organized as follows: Sec.~\ref{sec:physics} introduces the elliptic PDEs governing black hole perturbations and elastodynamics, Sec.~\ref{sec:linear} describes the resulting sparse linear systems and traditional preconditioners, Sec.~\ref{sec:gnn} details our GNN-based neural preconditioner and new training+loss strategy, Sec.~\ref{sec:results} presents our results, and Sec.~\ref{sec:conclusions} summarizes our conclusions.

\section{Physics overview}
\label{sec:physics}

\subsection{Elliptic PDEs}

Consider the field $u$ to be a function of two coordinates, $x$ and $y$, that satisfies a linear second-order PDE involving $x$ and $y$ derivatives
\begin{align}
\big( C_{xx} \partial^2_x + C_{xy} \partial_x \partial_y + C_{yy} \partial^2_y + C_x\partial_x&+C_y\partial_y+C_0 \big) u(x,y) \notag
\\& = f(x,y) \, , \label{eq:pde}
\end{align}
where the PDE coefficients, $C_*$, are generally functions of $x$ and $y$, and $f(x,y)$ is the source. Equation~\eqref{eq:pde} is classified as elliptic when $C_{xy}^2-4C_{xx}C_{yy}<0$, which is the classification of interest to this project. Elliptic PDEs involve a fixed global solution that is typically uniquely determined by imposing boundary conditions. Naturally, appropriate choices for $C_*$, $f$, and each boundary condition are motivated by the associated physics context. Common boundary conditions include Dirichlet: $u|_\text{bound} = c$, Neumann: $\hat{n} \cdot \vec{\nabla}u|_\text{bound} = c$ (where $\hat{n}$ is normal to the boundary and $\vec{\nabla}$ is the gradient), and Robin: $(\hat{n} \cdot \vec{\nabla} + b)u|_\text{bound} = c$. 

Discretizations of elliptic PDEs produce sparse linear systems. Arguably the simplest PDE discretization appropriate for this investigation is the second-order finite-difference method
\begin{align}\label{eq:fd_stencil}
\partial^2_x \, u(x,y) \simeq \frac{u(x+\Delta x,y)-2u(x,y)+u(x-\Delta x,y)}{\Delta x^2} \, ,
\end{align}
where $\Delta x$ is the $x$ grid spacing (and similar for $\Delta y$).  For simplicity, we adopt this straightforward discretization, although our intended downstream applications use more sophisticated spectral and discontinuous Galerkin methods~\cite{Vu_2026}. Other derivatives are approximated by similar logic. The scenarios considered in this work involve a two-dimensional rectangular numerical domain with a constant grid spacing along each direction (typically with two different spacings for the two directions). Through this logic, each field value at each position contributes an unknown that is determined via the solution vector of the linear algebra problem.

In \cref{sec:intro} we discussed many important physical systems that are governed by elliptic PDEs, ranging from simple electrostatics to ambitious nanophotonic inverse design~\cite{Molesky_2018,Shin_2013}. This project primarily focuses on the elliptic PDEs encountered when modeling the spacetime metric of asymmetric mass compact binaries by solving the perturbed Einstein equations via $m$-modes in the frequency domain, which is important for accurate gravitational wave predictions. In addition, we investigate a cross-disciplinary application involving elastodynamics with driven waves near the surface of a solid, which has engineering and materials science applications, but that we introduce in the context of geophysics and seismology.

\subsection{Gravitational waves from asymmetric binaries}

\subsubsection{Black hole perturbation theory}

\begin{figure}
\centering
\includegraphics[width=0.7\columnwidth]{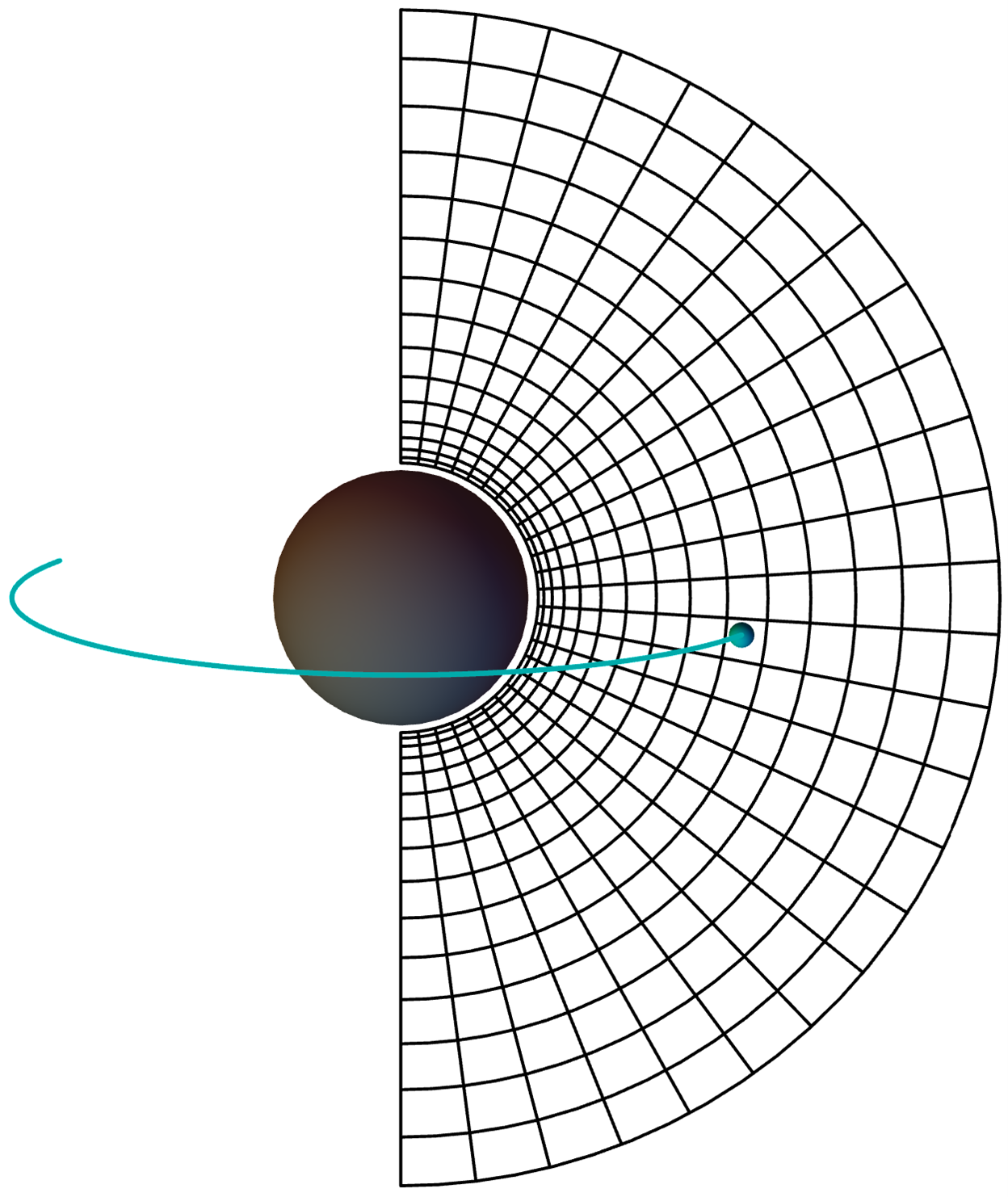}
\vspace{1pt} 
\includegraphics[trim={0 3 0 0},clip,width=\columnwidth]{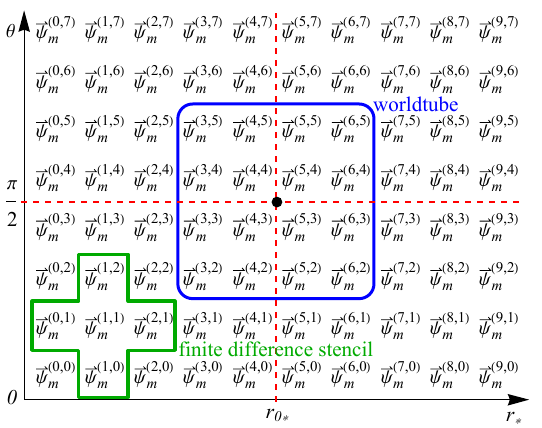}
\caption{Top: a small compact object (teal) on a circular orbit around a larger black hole is depicted alongside the $(r_*,\theta)$ grid. Notice how the radial tortoise coordinate, $r_*$, resolves the near-horizon region. Separation of the $\varphi$ dependence into $m$-modes produces a 2D spatial domain. Bottom: the same type of 2D grid mapped to a rectangular numerical domain with constant spacings $\Delta r_*$ and $\Delta\theta$, where the ten-component field $\vec{\psi}_m^{\,(i,j)}$ is stored at each grid point $(r_{*i},\theta_j)$. The red dashed lines at $r_*=r_{0*}$ and $\theta=\pi/2$ intersect at the position of the smaller body (black dot), and the blue worldtube is the effective source region. The green cross shows the five-point stencil of the second-order finite-difference discretization, which couples each grid point to its nearest neighbors in $r_*$ and $\theta$. Only a few grid points are shown for clarity.}
\label{fig:discretization}
\end{figure}

Consider a black hole or neutron star with mass~$m_0$ in a relativistic bound orbit around a larger black hole with mass~$M$, see Fig.~\ref{fig:discretization}. The mass-ratio is $q\equiv M/m_0$. We describe the dynamics of this asymmetric-mass system by considering the larger black hole to be the dominant contribution to the gravitational field with the smaller body contributing small perturbations
\begin{align}
g_{\alpha\beta} = g^\text{BH}_{\alpha\beta} + \frac{1}{q} \, h_{\alpha\beta} + \mathcal{O}\left( \frac{1}{q^2} \right) \, ,
\end{align}
where $g_{\alpha\beta}$ is the spacetime metric of general relativity, $g^\text{BH}_{\alpha\beta}$ is the dominant contribution from the larger black hole, which is known exactly, and $h_{\alpha\beta}$ is the metric perturbation that encodes two-body interactions, which must be calculated numerically. The metric is a symmetric tensor field where each of the ten components are functions of the four spacetime coordinates, ($t$, $r_*$, $\theta$, $\varphi$), where $r_*$ is the tortoise coordinate~\cite{Regge1957}. This approach is called black hole perturbation theory: the asymptotic metric perturbation describes the gravitational waves and the local metric perturbation's backreaction on the orbital motion of the smaller body is the gravitational self-force~\cite{Mino_1997,Quinn_1997}.

According to general relativity, the spacetime metric satisfies the Einstein field equations
\begin{align}
G_{\alpha\beta} = 8\pi\,T_{\alpha\beta} \, , \label{eq:efe}
\end{align}
where $G_{\alpha\beta}$ is the Einstein tensor, which is a second-order non-linear differential operator acting on $g_{\alpha\beta}$, and $T_{\alpha\beta}$ is the stress-energy tensor. This produces a system of ten coupled non-linear PDEs governing the ten components of $g_{\alpha\beta}$. Because $g^\text{BH}_{\alpha\beta}$ is an exact solution of Eq.~\eqref{eq:efe}, the leading non-trivial term in the field equation expansion governs $h_{\alpha\beta}$. Furthermore, at the first perturbative order the non-linear terms are negligible.

\subsubsection{Lorenz gauge field equations with $m$-modes}

General relativity involves four degrees of gauge freedom, and we adopt the reasonably well-behaved and commonly used Lorenz gauge
\begin{align}
\nabla^\beta \bar{h}_{\alpha\beta} = 0 \, ,
\end{align}
where $\nabla^\beta$ is a covariant derivative compatible with $g^\text{BH}_{\alpha\beta}$ and $\bar{h}_{\alpha\beta}$ is the trace-reversed metric perturbation: $\bar{h}_{\alpha\beta}=h_{\alpha\beta}-g^\text{BH}_{\alpha\beta} \, g_\text{BH}^{\mu\nu}\, h_{\mu\nu}/2$. In the Lorenz gauge, the first-order perturbation of Eq.~\eqref{eq:efe} takes the following form
\begin{align}
\nabla^\alpha \nabla_\alpha \,\bar{h}_{\mu\nu}+2R^{\alpha\;\beta}_{\;\,\mu\;\nu}\,\bar{h}_{\alpha\beta}= -16\pi \, T_{\mu\nu} \, , \label{eq:lgfe}
\end{align}
where $R^{\alpha\;\beta}_{\;\,\mu\;\nu}$ is the Riemann tensor of $g^\text{BH}_{\alpha\beta}$. Thorough reviews of black hole perturbation theory and self-force concepts are available in the literature~\cite{Poisson:2011nh,Barack:2018yvs,Pound_2021}.

Multiple techniques to solve Eq.~\eqref{eq:lgfe} have been investigated. Although spinning black hole backgrounds are not spherically symmetric, certain aspects of the gravitational perturbation are accessible through the Teukolsky formalism, which involves full separation of variables and spin-weighted spheroidal harmonics~\cite{Teukolsky_1973}; however, starting at non-linear order (second-order) in perturbation theory, problematic mode-coupling~\cite{Miller_2016} and slow polar convergence introduces considerable obstacles to the full separation of variables approach for spinning black hole backgrounds. To enable a more direct path to higher-order perturbations, this work pursues the frequency domain $m$-mode approach~\cite{Osburn_2022,Macedo_2024,Vu_2026,osburn2026schwarzschildperturbationslorenzgauge,nishimura2026,leather2026b,cownden2026,casals2026}, where only the $t$ and $\varphi$ variables are separated
\begin{align}
\label{eq:modes}
\bar{h}_{\mu\nu}(t,r_*,\theta,\varphi) = \sum_m \bar{h}^m_{\mu\nu}(r_*,\theta)\, e^{im(\varphi-\Omega t)}\, ,
\end{align}
where $\bar{h}^m_{\mu\nu}$ is an $m$-mode of the metric perturbation and $\Omega$ is the angular velocity for a smaller body in circular motion, see Fig.~\ref{fig:discretization}. For simplicity, we vectorize the ten independent components of $\bar{h}^m_{\mu\nu}$
\begin{align}
\label{eq:fieldVec}
\vec{\psi}_m \simeq (\bar{h}^m_{tt},\bar{h}^m_{tr},\bar{h}^m_{t\theta},\bar{h}^m_{t\varphi},\bar{h}^m_{rr},\bar{h}^m_{r\theta},\bar{h}^m_{r\varphi},\bar{h}^m_{\theta\theta},\bar{h}^m_{\theta\varphi},\bar{h}^m_{\varphi\varphi}) \, ,
\end{align}
where scale factors are applied to each component to achieve better numerical behavior at the boundaries, see Ref.~\cite{osburn2026schwarzschildperturbationslorenzgauge} for full details (which is a frequency domain extension of Ref.~\cite{Dolan_2013}). Although we focus here on perturbations of non-spinning (Schwarzschild) black holes, we recently achieved generalization to spinning (Kerr) black hole backgrounds~\cite{nishimura2026}. 

\begin{figure}
\includegraphics[trim={0 3 0 0},clip,width=\columnwidth]{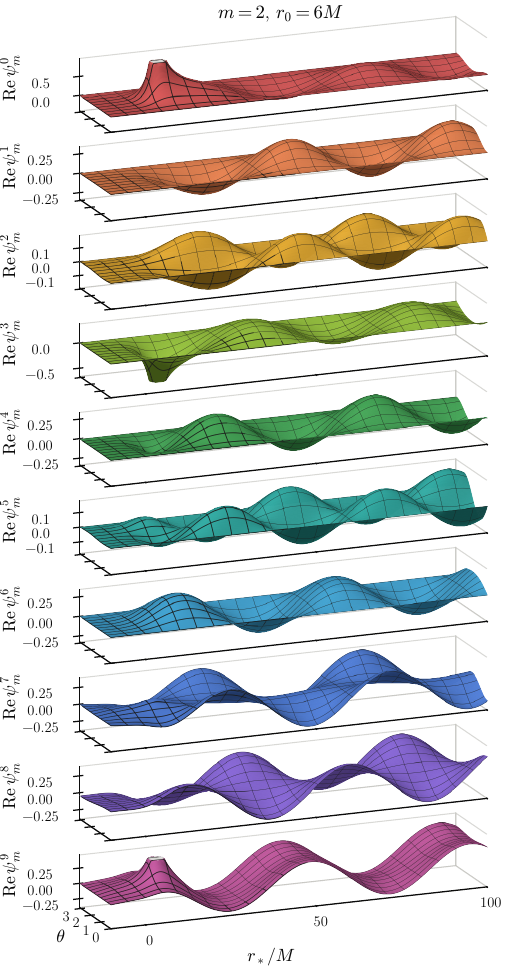}
\caption{The ten components of the $m=2$ spacetime metric perturbation, $\vec{\psi}_m \propto \bar{h}^m_{\alpha\beta}$, for a small body in circular motion with $r_0=6M$, obtained by solving \cref{eq:PDEs} on the $(r_*,\theta)$ domain of \cref{fig:discretization}. From top to bottom, ${\rm Re}\,\psi_m^i$ follows the ordering of \cref{eq:fieldVec} with scale factors from Ref.~\cite{osburn2026schwarzschildperturbationslorenzgauge}. The divergent features near the small body ($r_*\simeq r_{0*}$, $\theta\simeq\pi/2$) show the expected Coulomb-type near-field behavior, while the oscillations at large $r_*$ are the outgoing gravitational waves.} \label{fig:fields} 
\end{figure}

Substituting Eq.~\eqref{eq:modes} into Eq.~\eqref{eq:lgfe} produces a system of ten coupled linear elliptic PDEs governing each $m$-mode
\begin{align}
\label{eq:PDEs}
\left(\partial^2_{r_*} + p\,\partial^2_\theta  + A_m \partial_{r_*}  + B_m \partial_\theta  + C_m \right) \vec{\psi}_m = \vec{S}_m \, ,
\end{align}
where $\vec{S}_m$ is the $m$-mode source from the smaller body, $p$ is a known positive function of $r_*$, and $A_m$, $B_m$, and $C_m$ are 10$\times$10 matrices where each element is a known function of $r_*$ and $\theta$. We discretize the $r_*$-$\theta$ domain according to a grid with fixed $\Delta r_*$ and $\Delta\theta$, which is a rectangular numerical domain, see Fig.~\ref{fig:discretization}. At the polar boundaries we impose a mixture of Dirichlet and Neumann boundary conditions to ensure $\bar{h}_{\mu\nu}$ is regular. At the outer $r_*$ boundary we impose medium-order radiation boundary conditions involving $\partial_{r_*}^3\vec{\psi}_m$ (and lower derivatives); at the inner $r_*$ boundary near the larger black hole's event horizon we impose a mixture of gauge conditions and high-order radiation boundary conditions involving $\partial_{r_*}^5\vec{\psi}_m$ (and lower derivatives) to handle otherwise problematic behavior there, see Ref.~\cite{osburn2026schwarzschildperturbationslorenzgauge} for full details. Figure~\ref{fig:fields} previews a numerical solution of Eq.~\eqref{eq:PDEs} with appropriate boundary conditions, which displays the asymptotic ($r_*\rightarrow\infty$) behavior of the gravitational waves.

\subsection{Elastodynamics application}
\label{sec:elasto}

\begin{figure}
\includegraphics[width=\columnwidth]{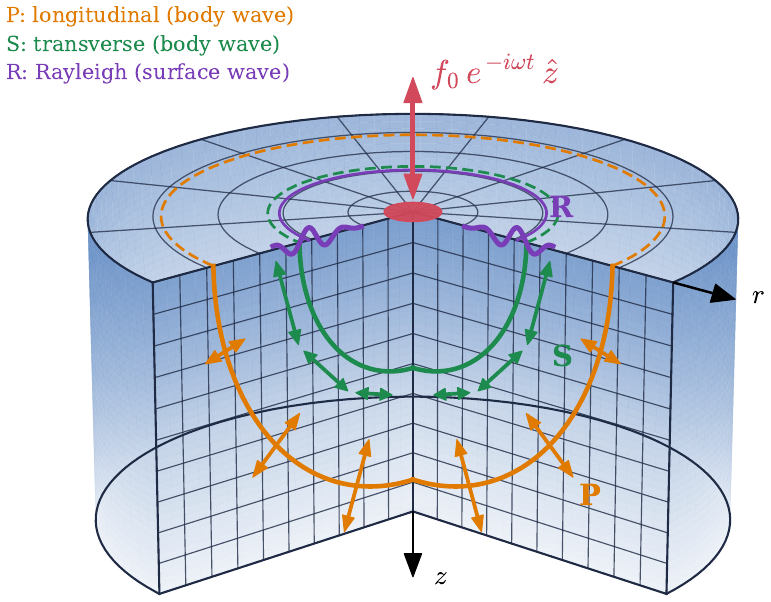}
\caption{Diagram illustrating our elastodynamics application. A cutaway of the $z>0$ elastic half-space shows the axisymmetric $(r,z)$ numerical grid. An oscillating Gaussian pressure distribution with total force $f_0\,e^{-i\omega t}\,\hat{z}$ (red) is applied to the $z=0$ surface, centered on $r=0$, which launches three types of elastic waves: the longitudinal P wave (orange), the slower transverse S wave (green), and the Rayleigh surface wave (purple). Although elastodynamics involves numerous applications, this specific scenario appears in certain theoretical seismology calculations. Double-headed arrows indicate the direction of oscillations according to their respective polarizations, and the dashed circles mark where the body-wave fronts meet the surface.}
\label{fig:halfspace} 
\end{figure}

Consider a linear isotropic non-homogeneous elastic solid with density $\rho$ and Lam\'e parameters $\lambda$ and $\mu$. The non-homogeneity implies potential position-dependence of these parameters. We assume the presence of a driving force with sinusoidal time dependence, $\vec{f}\,e^{-i\omega t}$, which will produce a displacement vector field that also oscillates harmonically: $\vec{u}\,e^{-i\omega t}$. Newton's second law applied to a volume element provides the equation of motion
\begin{align}
\label{eq:newt}
- \rho\, \omega^2 \vec{u} = \vec{f} + \vec{F} \, ,
\end{align}
where $\vec{f}$ is an externally applied force density and $\vec{F}$ is the restoring force density that follows from the divergence of the stress tensor~\cite{Aki_2002}
\begin{align}
\label{eq:F}
\vec F = (\lambda+\mu)\,\vec\nabla(\vec\nabla\cdot\vec u) + & \mu\,\nabla^2\vec u + (\vec\nabla\cdot\vec u)\,\vec\nabla\lambda 
\\ & + 2\,(\vec\nabla\mu\cdot\vec\nabla)\,\vec u
  + \vec\nabla\mu\times(\vec\nabla\times\vec u)\, . \notag
\end{align}
The first two terms in the restoring force produce the familiar Navier operator while the remaining terms account for the non-uniformity of the solid.

We introduce the following simplifying assumptions: the elastic solid fills only the $z>0$ half-space, $\rho$ is uniform, $\mu$ and $\lambda$ are linear functions of $z$ such that the ratio $\lambda /\mu$ is uniform, and $\vec{f}$ vanishes everywhere except for an axisymmetric Gaussian pressure distribution at the $z=0$ boundary
\begin{align}
&\mu = \mu_0 \left( 1 + \frac{z}{z_0} \right) \, ,
\\
&\lambda = \kappa_0 \, \mu,
\\
&\vec{f} = \frac{f_0}{\pi b^2} e^{-r^2/b^2} \delta(z) \, \hat{z} \, ,
\end{align}
where $z_0$ is the depth at which $\mu$ and $\lambda$ become double their surface values, $\kappa_0$ is related to the Poisson ratio, $\nu$, by $\kappa_0 = 2\nu/(1-2\nu)$, $f_0$ is the total force exerted on the surface, $r$ is the cylindrical radial coordinate, $b$ is the Gaussian radius, $\delta(z)$ is the Dirac delta function, and $\hat{z}$ is a unit vector. Naturally, $\rho$, $\mu$ and $\lambda$ vanish when $z<0$. This produces a simple but reasonable model for certain seismic waves near the surface, see Fig.~\ref{fig:halfspace}. The limit $z_0\to\infty$ is the homogeneous half-space Lamb's problem~\cite{Lamb_1904}, and the simultaneous limit $\mu_0\to 0$ and $z_0\to 0$ with fixed $\mu_0/z_0$ is the Gibson's soil scenario~\cite{Gibson_1967} where $\mu\propto z$.

It is convenient to adopt dimensionless units; moving forward, we measure stress in units of $\mu_0$ and density in units of $\rho$, which produces a unit surface shear speed $c_0 = \sqrt{\mu_0/\rho} = 1$; we measure length in units of the surface shear wavelength $\Lambda = 2\pi c_0/\omega$, which implies $\omega = 2\pi$. The homogeneous operator then depends on two dimensionless parameters only: $z_0$ and $\kappa_0$. Although $\omega$ naively seems like it should be a controllable external parameter, in this formulation we would implement something like an effective decrease in wavelength due to an increase of source frequency by expanding the size of the domain (and source) to admit additional wavelength spans.

These assumptions lead to an axisymmetric $\vec{u}$, which we describe in dimensionless units and cylindrical coordinates according to Eqs.~\eqref{eq:newt} and~\eqref{eq:F}
\begin{align}
\label{eq:explicit-r}
& 0 = (\kappa_0+2)\big(\partial_r^2 u_r + \frac{1}{r}\partial_r u_r - \frac{u_r}{r^2}\big)  + (\kappa_0+1)\partial_r\partial_z u_z \notag
\\&\qquad\;\;\;\; + \partial_z^2 u_r + \frac{1}{\mu z_0}(\partial_z u_r + \partial_r u_z) + \frac{4\pi^2}{\mu} u_r ,
\\&0 = (\kappa_0+2)\partial_z^2 u_z + (\kappa_0+1) \big( \partial_r\partial_z u_r + \frac{1}{r}\partial_z u_r \big) + \frac{1}{r}\partial_r u_z \notag
\\& + \partial_r^2 u_z+ \frac{1}{\mu z_0}\big( \kappa_0(\partial_r u_r + \frac{u_r}{r}) + (\kappa_0+2)\partial_z u_z\big) + \frac{4\pi^2}{\mu} u_z,
\label{eq:explicit-z}
\end{align}
where $u_r$ and $u_z$ are the non-vanishing cylindrical components of $\vec{u}$, and the source-free form of the equations is a consequence of our driving force being confined to the surface where it will be applied as part of a boundary condition. Equations~\eqref{eq:explicit-r} and \eqref{eq:explicit-z} form a system of two coupled linear elliptic PDEs.

The $z=0$ boundary conditions follow from the $rz$ stress tensor component vanishing and the $zz$ stress tensor component being prescribed by the surface pressure
\begin{align}
& \left( \partial_z u_r + \partial_r u_z \right)\big|_{z=0} = 0 \, ,
\\
& \Big( \kappa_0\big(\partial_r u_r + \frac{u_r}{r}\big) + (\kappa_0+2)\,\partial_z u_z \Big) \Big|_{z=0} = -\frac{f_0}{\pi b^2}e^{-r^2/b^2} \, .
\end{align}
The $r=0$ boundary conditions are motivated by preservation of regularity in the vicinity of the $r=0$ coordinate singularity
\begin{align}
& u_r \big|_{r=0} = 0 \, ,
\\
& \partial_r u_z \big|_{r=0} = 0 \, .
\end{align}
The two outer boundaries are located at $r_\text{max}$ and $z_\text{max}$; the associated boundary conditions are designed to absorb both the longitudinal P~wave and the transverse S~wave~\cite{Bayliss_1980,Higdon_1991}
\begin{align}
\Big(\hat{n}\cdot \vec{\nabla} -i k_P + \frac{1}{R} \Big)\Big( \hat{n}\cdot \vec{\nabla} -i k_S + \frac{1}{R} \Big) \, \vec{u} \, \big|_\text{outer} = 0 \, ,
\end{align}
where $\hat{n}$ is a unit vector normal to the boundary, $R$ is the spherical radius $R=\sqrt{r^2+z^2}$, and $k_{P/S}$ are the position-dependent wavenumbers of the P and S waves
\begin{align}
k_S = \frac{2\pi}{\sqrt{1+z/z_0}} \, , \qquad\;\;\;\; k_P = \frac{k_S}{\sqrt{2+\kappa_0}} \, .
\end{align}
$R$ and $k_{P/S}$ are all evaluated as constants at their respective boundary position prior to applying the boundary condition derivatives.

\section{Sparse linear systems}
\label{sec:linear}

\begin{figure*}
\includegraphics[width=\textwidth]{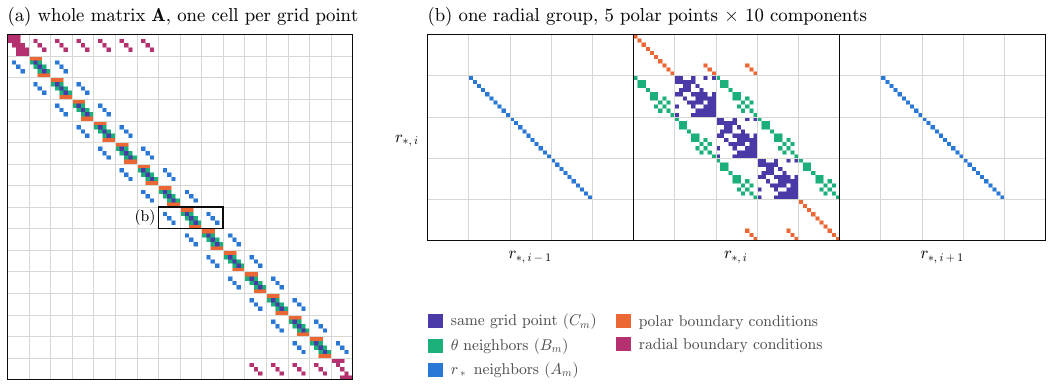}
\caption{Sparsity pattern of the matrix $\mathbf{A}$ for the black hole perturbation problem on a small illustrative grid ($16$ radial $\times$ $5$ polar grid points, 10 tensor components each, $800$ total unknowns). Colors indicate the term of the elliptic equations, \cref{eq:PDEs}, that produces each matrix element. Panel (a) shows the full matrix with one colored cell per grid point (condensing the $10\times 10$ tensor-component couplings into one colored cell) and with light gray boxes indicating a radial group of $5$ polar points each. The $5$-point stencil of \cref{fig:discretization} produces radial (light blue) and polar (green) couplings between neighboring grid points. The top and bottom strips indicate the radial boundary conditions (magenta), with a fifth radial derivative ($7$ points) at the inner boundary (top row) and a third radial derivative ($5$ points) at the outer boundary (bottom row). These radial boundary cells are eliminated from the system using the Schur complement, see Appendix~\ref{sec:schur}. Panel (b) shows one radial group with the $10\times 10$ tensor-component couplings expanded into individual colored cells. Light gray boxes indicate individual grid points in the polar direction. Tensor components are coupled pointwise at each grid point (dark blue diagonal) and between grid points through derivatives (light blue and green). The top and bottom rows of each radial group impose the polar boundary conditions (orange). For full details, see Ref.~\cite{osburn2026schwarzschildperturbationslorenzgauge}.}
\label{fig:matrix} 
\end{figure*}

Once the elliptic PDEs are discretized, such as with the finite difference method of \cref{eq:fd_stencil}, they form a linear system of equations,
\begin{equation}
    \mathbf{A} \mathbf{u} = \mathbf{b}
    \text{,}
\end{equation}
where $\mathbf{u}$ is the state vector of all unknowns (the value of the solution fields at all grid points in the finite difference method), $\mathbf{A}$ is a matrix that couples the unknowns (across grid points via derivatives, and across degrees of freedom per grid point via the elliptic equations), and $\mathbf{b}$ is the fixed source vector (the value of $\vec{S}_m$ at the grid points, or the effect of $\vec{f}$ imposed at the boundary).
\Cref{fig:matrix} presents a matrix $\mathbf{A}$ stemming from the second-order finite-difference discretization of the black-hole perturbation problem. The matrix is sparse, as the finite-difference stencil involves only a few neighboring grid points in the computation of derivatives. Other discretization methods can lead to different sparsity patterns: multi-domain spectral methods and discontinuous Galerkin methods typically lead to block-diagonal sparsity patterns with dense coupling within each spectral element and sparse coupling across elements (see \ccite{Vu_2026} for an example).

To solve the discretized equations, one must effectively invert the matrix~$\mathbf{A}$ to obtain the state vector $\mathbf{u}=\mathbf{A}^{-1}\mathbf{b}$. Note that solving the discretized equations even to machine precision does not mean that the continuum equations are solved exactly; a discretization error remains, $|\mathbf{u}-\mathbf{u}_\mathrm{exact}|$, that originates from representing the continuum equations with finite degrees of freedom, such as evaluating them on a grid. This discretization error is discussed in prior work~\cite{osburn2026schwarzschildperturbationslorenzgauge}, but is not the focus of this article. Instead, our focus is accurately and efficiently solving the discretized equations such that the residual error, $|\mathbf{A}\mathbf{u}-\mathbf{b}|$, is below a prescribed relative tolerance ($10^{-10}$ for this work).

Direct elimination methods to invert $\mathbf{A}$ quickly become infeasible at high resolution, in two or three dimensions, and for coupled multi-component equations. Instead, Krylov-type algorithms such as Conjugate Gradients (CG) and the Generalized Minimal Residual method (GMRES) are common choices in computational physics. They solve the linear system iteratively, starting from an initial guess, up to the prescribed residual error tolerance. In addition, they require only the matrix-vector product, $\mathbf{A}\mathbf{u}$, which can often be evaluated efficiently on distributed-memory computers by taking the underlying discretization and the resulting sparsity of the linear operator into account.

The convergence rate of a Krylov-type algorithm can be greatly enhanced by a preconditioner, $\mathbf{P}$, which approximates the matrix $\mathbf{A}$ but is easier to invert. The preconditioned system,
\begin{align}
    &\mathbf{P}^{-1}\mathbf{A}\mathbf{u}=\mathbf{P}^{-1}\mathbf{b}
    &\quad \text{(left-preconditioning)}\\
    \text{or}\quad
    &\mathbf{A}\mathbf{P}^{-1}\mathbf{v}=\mathbf{b}, \quad \mathbf{u}\equiv\mathbf{P}^{-1}\mathbf{v}
    &\quad \text{(right-preconditioning)}
\end{align}
is easier to solve by the Krylov-type algorithm when the preconditioned operator,  $\mathbf{P}^{-1}\mathbf{A}$ or $\mathbf{A}\mathbf{P}^{-1}$, has favorable properties, such as a condition number closer to one (see \ccite{saad2003iterative} for an overview). The choice of preconditioner is typically very problem-dependent, as it can involve knowledge of the solution structure (mode content, physical approximations), the domain geometry (refinement, resolution levels), or the computer architecture (parallelizable subproblems, hardware-aware algorithms). ``Flexible'' variants of Krylov-type algorithms (FCG, FGMRES) allow $\mathbf{P}$ to change in each iteration, allowing a wide range of preconditioning techniques. Common techniques are variants of multigrid methods (solving on coarser grids for the large-scale modes in the solution) and Schwarz methods (splitting the problem into subproblems that can be solved in sequence or in parallel), and these techniques can often be combined to leverage their individual strengths.

ML-based preconditioners have great potential because producing fast approximate solutions is a strength of machine learning. Even when they are not sufficiently accurate to solve a demanding physics problem by themselves, ML preconditioners can complement a traditional Krylov-type algorithm to accelerate its convergence without sacrificing accuracy because the Krylov framework guarantees a prescribed maximum residual error. This is a realization of the principle depicted in \cref{fig:idea}, where the Krylov-type algorithm enforces correctness and the ML preconditioner acts as an accelerator.

The elliptic equations considered in this work are particularly challenging, as they are of Helmholtz-type, schematically: $(\partial_x^2 + \omega^2)u(x)=0$. Phrased in physics terms, preconditioning Helmholtz-type problems is especially challenging because local wave content in the solution must propagate across the entire domain and boundary conditions through iterations of the solver. Local solution methods are therefore at a disadvantage. To mitigate these challenges, various techniques have been developed. A widely used technique is a complex shift applied to the preconditioner, for example by replacing $\omega^2$ with $(1+i\beta)\,\omega^2$. It moves the eigenvalue spectrum off the real axis into the complex plane and damps the spurious oscillations amplified by local solvers, at the cost of making the preconditioner a less accurate approximation to the original matrix~\cite{Erlangga2006,Gander_2019}. Deflation techniques project known modes out of the preconditioned solve, and optimized Schwarz methods take the mode content into account when coupling subdomains~\cite{Gander_2019}. Despite these developments, preconditioning Helmholtz-type problems remains an open problem in numerical analysis~\cite{ErnstGander2012,Gander_2019}.

In the following we first describe the traditional methods used in this work: GMRES with deflated restarts and Incomplete LU (ILU) factorization, before describing our graph neural network (GNN) preconditioner approach in detail in \cref{sec:gnn}.

\subsection{GMRES with deflated restarts}
\label{sec:gmres}

We use a complex-valued, right-preconditioned GMRES algorithm with deflated restarts (DR)~\cite{morgan2002gmres,parks2006recycling} for the outer Krylov-type iterations. It minimizes the residual error $\|\mathbf{b}-\mathbf{A}\mathbf{u}\|$ and stops when it reaches a relative tolerance of $10^{-10}$. Restarts avoid the growing cost of orthogonalization against all prior basis vectors, but may also slow convergence because they discard the basis vectors that the previous cycle just resolved. Deflation mitigates this by keeping the slowest converging directions from the previous cycle and orthogonalizing against them in the next cycle. We restart every 50 iterations and keep 10 deflated directions. These deflated directions turn out to be the near-resonant modes of the Helmholtz-type operator, which are the most difficult to resolve.

\subsection{Incomplete LU preconditioner}

As a baseline preconditioner we use an Incomplete LU (ILU) factorization of the discretized operator.
The LU factorization is a standard technique to solve linear systems, where the matrix $\mathbf{A}$ is factored into a lower triangular matrix $\mathbf{L}$ and an upper triangular matrix $\mathbf{U}$ such that $\mathbf{A} = \mathbf{L}\mathbf{U}$. The solution can then be obtained by first solving $\mathbf{L}\mathbf{y} = \mathbf{b}$ for $\mathbf{y}$ and then solving $\mathbf{U}\mathbf{u} = \mathbf{y}$ for $\mathbf{u}$. However, for large sparse matrices, the full LU factorization can introduce significant fill-in, leading to increased memory usage and computational cost.
The ILU factorization is a sparse approximation to the full LU factorization, where fill-in outside a prescribed sparsity pattern is dropped. We use the level-of-fill approach~\cite{saad2003iterative}, where ILU(0) keeps only the original sparsity pattern and ILU(p) keeps fill-in of level $\leq p$, as defined by \ccite{saad2003iterative}.
Therefore, the preconditioner $\mathbf{P}_\text{ILU(p)}$ is given by
\begin{align}
    \mathbf{P}_\text{ILU(p)} = \mathbf{L}_p \mathbf{U}_p = \mathbf{A} + \mathbf{R}_p \, ,
\end{align}
where $\mathbf{R}_p$ is the dropped fill-in.

It is important to note that ILU is a local preconditioner, meaning that each factor entry depends only on nearby matrix entries. This locality makes ILU excellent for short wavelengths, as it can effectively capture the local behavior of the operator. However, for long wavelengths, the local nature of ILU can lead to poor approximations of the operator. This is particularly problematic for Helmholtz-type problems, where the operator has near-resonant modes that are not well captured by local preconditioners.

Note also that the accuracy of the factorization and its usefulness as a GMRES preconditioner are different things. A more accurate factorization does not necessarily lead to better convergence of GMRES, as the preconditioner must also effectively capture the relevant modes of the operator. In the next section we will construct a neural preconditioner that mirrors ILU(0) in fill pattern and application cost, but is trained on the GMRES residuals to learn an approximation of the operator that leads to more effective preconditioning.

\section{Neural network preconditioner}
\label{sec:gnn}

\begin{figure*}[t]
\includegraphics[trim={5 2 6 0},clip,width=\textwidth]{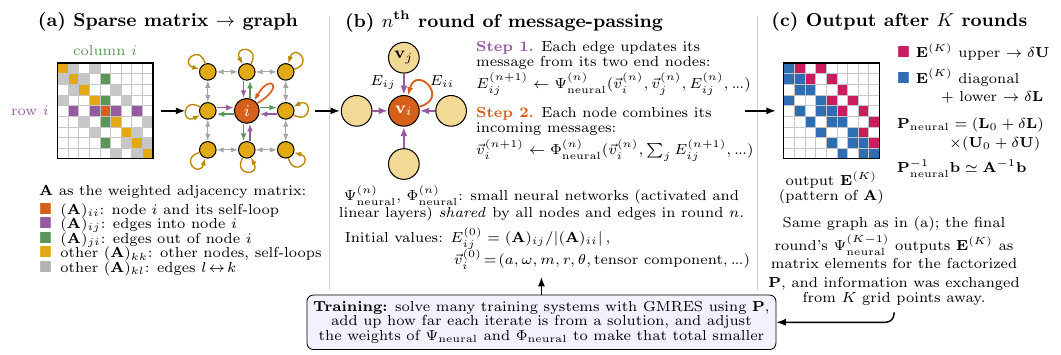}
\caption{Graph neural network (GNN) construction of the neural preconditioner. (a) The sparse matrix $\mathbf{A}$ is interpreted as the weighted adjacency matrix of a directed graph, illustrated here for a $3\times3$ grid with a (at most) five-point stencil and one unknown per grid point. Each diagonal element $(\mathbf{A})_{ii}$ is a node with a self-loop (orange for node $i$), and the off-diagonal elements in row $i$ (column $i$) are edges into (out of) node $i$, shown in purple (green). (b) In the $n^{\rm th}$ message-passing round, each edge feature $E_{ij}^{(n)}$ is first updated from the features of its two nodes by the neural network $\Psi^{(n)}_{\rm neural}$ (\cref{eq:Psi}), and each node feature $\vec{v}^{(n)}_i$ is then updated by $\Phi^{(n)}_{\rm neural}$ from its connected edge features (\cref{eq:Phi}). Note that our $\Phi^{(n)}_{\rm neural}$ uses mean aggregation according to \cref{eq:Phi}, which this figure's truncated presentation omits. The same $\Psi^{(n)}_{\rm neural}$ and $\Phi^{(n)}_{\rm neural}$ are shared by all edges and nodes within a round. The initial edge features are the diagonally scaled matrix elements (\cref{eq:initEdge}), and the initial node features encode the physical context of each unknown (\cref{eq:initNode}). (c) After $K$ rounds ($K=4$ in this work), the final edge features $E^{(K)}_{ij}$ produce $\mathbf{P}_\text{neural}$: the diagonal and lower-triangular elements (blue) produce $\mathbf{\delta L}$ and the upper-triangular elements (magenta) produce $\delta\mathbf{U}$, which correct the ILU(0) factors to form $\mathbf{P}_{\rm neural}$ (\cref{eq:Pneural,eq:dLdiag,eq:dL,eq:dU}). Because each round exchanges information between neighboring grid points, each element of $\mathbf{P}_{\rm neural}$ depends on matrix elements up to $K$ grid points away. During training (bottom box), GMRES solves of the training systems using $\mathbf{P}_{\rm neural}$ provide the loss function of \cref{eq:loss}, whose gradient is used to update the weights of $\Psi^{(n)}_{\rm neural}$ and $\Phi^{(n)}_{\rm neural}$.}
\label{fig:architecture} 
\end{figure*}

\subsection{Graph neural network and adjacency matrices}

One common and useful representation of a graph is via its adjacency matrix. Each diagonal matrix position represents a node, and off-diagonal matrix elements represent edges. Undirected graphs have symmetric adjacency matrices, while directed graphs are represented by non-symmetric adjacency matrices. Often features are assigned to each edge of a graph, and for edge features that are simple numbers it is customary for non-zero elements of the adjacency matrix to represent edge features, which is a weighted adjacency matrix. Finally, for graphs with self-loop edges that connect a node to itself, self-loop features are represented by diagonal matrix element values.

By that logic, an arbitrary square matrix can be interpreted as the weighted adjacency matrix of a directed graph with self-loops and numerical edge features, see Fig.~\ref{fig:architecture}(a). We adopt the weighted adjacency matrix interpretation for our sparse matrix $\mathbf{A}$, which enables us to use $\mathbf{A}$ as an input for a graph neural network (GNN)~\cite{GNN1,GNN2}, and similarly to extract a preconditioner, $\mathbf{P}_\text{neural}$, as the GNN output. We build $\mathbf{P}_\text{neural}$ as a pre-factorized matrix to enable the fast-inverse feature required for effective preconditioning. Our overall strategy, which assumes each factorized preconditioner matrix element can be calculated effectively using information from nearby nodes and edges, is an extension of the learned factorization approach~\cite{hausner2024,hausner2024GMRES} with new improvements to significantly outperform traditional methods. We will provide full GNN details for the black hole perturbation theory case here, and then explain any deviations necessary for the cross-disciplinary elastodynamics application in Sec.~\ref{sec:elasto_accel}.

The GNN mechanism for transmitting information among nearby nodes and edges is called message passing. While initial edge features, $E^{(0)}_{ij}$, have a natural interpretation as matrix elements, we also attach features to each node. The initial node features, $\vec{v}^{\,(0)}_{i}$, are assigned as vectors that encode additional context about the associated diagonal matrix element, such as the spatial position and tensor component of the corresponding unknown field:
\begin{align}
\label{eq:initEdge}
& E^{(0)}_{ij} = \frac{(\mathbf{A})_{ij}}{|(\mathbf{A})_{ii}|} \, ,
\\
\label{eq:initNode}
& \vec{v}^{\,(0)}_i = \big(\omega, m, r_{*i}, \theta_i, \vec{c}^\text{ tensor}_i, f^\text{bound}_{r_{*i}}, f^\text{bound}_{\theta_i}\big) ,
\end{align}
where $\vec{c}^\text{ tensor}_i$ is a ``one-hot'' vector where one of the ten vector components will be non-zero to indicate the tensor component of the corresponding unknown field, and $f^\text{bound}_{r_{*i}/\theta_i}$ are three-state flags indicating whether $r_{*i}$ or $\theta_i$ is at a boundary and, if so, which boundary. Note that a Kerr implementation would also include $a$ as an input to $\vec{v}^{\,(0)}_i$ (as depicted in Fig.~\ref{fig:architecture}(b)), but our Schwarzschild implementation naturally omits this. Then the $n^\text{th}$ message-passing round involves updating all edge and node features by inputting nearest neighbor values into fully-connected neural networks $\Psi^{(n)}_\text{neural}$ and $\Phi^{(n)}_\text{neural}$, see Fig.~\ref{fig:architecture}(b):
\begin{flalign}
\label{eq:Psi}
E^{(n+1)}_{ij} = \Psi^{(n)}_\text{neural}\big(\vec{v}^{\,(n)}_{i},\vec{v}^{\,(n)}_{j},E^{(n)}_{ij},E^{(n)}_{ji},\text{sign}(i-j), f^\text{tr}_{ij}\big)\,, &&
\\
\label{eq:Phi}
\vec{v}^{\,(n+1)}_{i} = \Phi^{(n)}_\text{neural}\big( \vec{v}^{\,(n)}_{i}, \tfrac{1}{N_\text{row}}\textstyle \sum_j E^{(n+1)}_{ij}, \tfrac{1}{N_\text{col}}\textstyle \sum_j E^{(n+1)}_{ji} \big)\, , &&
\end{flalign}
where $\text{sign}(i-j)$ is a three-state flag indicating whether the ($i$,$j$) matrix position is above, below, or on the diagonal, $f^\text{tr}_{ij}$ is a two-state flag that tells whether the transpose partner, $E^{(n)}_{ji}$, is non-zero (otherwise it is not part of the graph), and $N_\text{row/col}$ is the number of non-zero matrix elements in the corresponding row or column. This message-passing strategy where the mean of adjacent edge features is used to calculate the next node feature is called mean aggregation (note that Fig.~\ref{fig:architecture}(b) shows a simplified picture for brevity). 

After $K$ message-passing rounds, the output of $\Psi^{(K-1)}_\text{neural}$ produces $E^{(K)}_{ij}$, which is used to calculate the non-zero matrix elements of the factorized preconditioner $\mathbf{P}_\text{neural}$. Specifically, we train the neural networks to produce a correction to the ILU(0) preconditioner:
\begin{align}
&\mathbf{P}_\text{neural} = (\mathbf{L}_0+\mathbf{\delta L})(\mathbf{U}_0+\delta\mathbf{U}) \nonumber \\
\label{eq:Pneural}
&\phantom{\mathbf{P}_\text{neural}} \equiv \mathbf{A} + \delta\mathbf{A} + \mathbf{R}_0 \, ,
\\
\label{eq:dLdiag}
& (\mathbf{\delta L})_{ii} = E^{(K)}_{ii}\,|(\mathbf{A})_{ii}| \, ,
\\
\label{eq:dL}
& (\mathbf{\delta L})_{ij} = E^{(K)}_{i>j}\,|(\mathbf{L}_0)_{jj}| \, ,
\\
\label{eq:dU}
& (\mathbf{\delta U})_{ij} = E^{(K)}_{i<j} \, ,
\end{align}
where $\mathbf{\delta L}$ and $\delta\mathbf{U}$ are learned corrections to ILU(0), $\delta\mathbf{A}$ is an equivalent learned perturbation of $\mathbf{A}$ such that $\mathbf{P}_\text{neural}$ is an approximate factorization of $\mathbf{A}+\delta\mathbf{A}$, and zeros are implied for factorized matrix elements outside the appropriate lower/upper regions, see Fig.~\ref{fig:architecture}(c). Because the graph structure was preserved, $\mathbf{\delta L}$ has the same sparsity pattern as $\mathbf{L}_0$ and $\mathbf{\delta U}$ has the same sparsity pattern as $\mathbf{U}_0$ (except that $(\mathbf{\delta U})_{ii}\equiv 0$ considering that $(\mathbf{U}_0)_{ii}\equiv 1$). The scale factors confine $E^{(K)}_{ij}$ to a relatively small range of sizes, which aids training. Scaling the diagonal correction by $|(\mathbf{A})_{ii}|$ rather than by the ILU(0) pivot itself further allows the network to repair ILU near-zero pivots. Because the neural network produces a $\mathbf{P}_\text{neural}$ that is equivalent to an ILU(0) factorization of the modified operator $\mathbf{A} + \delta\mathbf{A}$, it is possible to reconstruct $\delta\mathbf{A}$ from the trained network (according to \cref{eq:Pneural}), where $\mathbf{R}_0$ contains the ILU dropped entries outside the sparsity pattern.

One powerful feature of this GNN strategy is that, within a certain message-passing round, all edges use identical weights and biases for $\Psi^{(n)}_\text{neural}$ and all nodes use identical weights and biases for $\Phi^{(n)}_\text{neural}$. This enables a compact memory footprint and comparatively faster training even when the dimensions of $\mathbf{A}$ become large. This weight-sharing feature is also used by transformers~\cite{Vaswani2017} and convolutional neural networks~\cite{lecun1989handwritten,Krizhevsky2012}; transformers are like a GNN with more global connectivity between nodes, while convolutional neural networks are localized similar to a typical GNN, but with a more rigid connectivity pattern. GNNs, transformers, and convolutional neural networks are all categorized within the geometric deep learning framework~\cite{Bronstein2017,Bronstein2021}.

\subsection{Training, hyperparameters, and inference}

Consider the vector $\vec{z}$ to represent the input for any example neural network; all of $\Psi^{(n)}_\text{neural}$ and $\Phi^{(n)}_\text{neural}$ are fully connected neural networks with two hidden layers:
\begin{align}
\Psi/\Phi^{(n)}_\text{neural}(\vec{z}\,) = W_3\,\sigma\Big(W_2\,\sigma\big(W_1 \vec{z}+\vec{b}_1\big)+ \vec{b}_2\Big)+\vec{b}_3 \, ,
\end{align}
where the output layer uses a linear activation function and the hidden layers use the sigmoid-weighted linear unit (SiLU) activation function;
\begin{align}
\sigma(x) = \frac{x}{1+e^{-x}} \, .
\end{align}
Note that, for the final message-passing round, we initialize $W_3$ and $\vec{b}_3$ with zeros prior to training so that the untrained network produces ILU(0) as a reasonable starting point. We find that $K=4$ message-passing rounds produces a good balance of speed and accuracy. In this second-order finite-difference context, that translates to exchanging information from four spatial grid-points away in both the $r_*$ and $\theta$ directions.

We implement these in \texttt{pytorch}, which does not support complex numbers; this affects $E^{(0)}_{ij}$ and $E^{(K)}_{ij}$, which are generally complex, so in reality we separate the real and imaginary parts as two different inputs/outputs for those networks. Furthermore, we vectorize $E^{(0<n<K)}_{ij}$ to have 32 components during intermediate message-passing rounds to enable deeper learning. Similarly, we increase the dimension of $\vec{v}^{\,(0<n<K)}_{i}$ to 32 during intermediate rounds (compared to 16 for $\vec{v}^{\,(0)}_{i}$), and all hidden layers have bias vectors with a dimension of 64. Because the final output is $E^{(K)}_{ij}$, we do not calculate $\vec{v}^{\,(K)}_{i}$ (which has no effect). These specifications imply the dimensions of each weight matrix and bias vector, and produce a total number of 86,978 GNN parameters. We use 32-bit floating-point numbers (floats) during training and then 16-bit floats for fast inference, which we find does not introduce a limiting factor in the iteration count or overall accuracy of the converged solution (64-bit numbers are used during key GMRES operations to preserve overall accuracy).

To train the network, we seek an appropriate loss function, $\mathcal{L}$, whose minimum corresponds to a high-performing $\mathbf{P}_\text{neural}$. Initially we tried the reference loss function candidates, which are stochastic estimates of the Frobenius norm $\big\|\mathbf{A}-\mathbf{P}\big\|^2$~\cite{hausner2024,hausner2024GMRES}. Unfortunately, those loss function candidates produced preconditioners that performed significantly worse than ILU(0) for our systems. Rather, we have adopted a loss function that directly measures how GMRES cycles reduce the residual according to a preconditioner produced by present network parameters:
\begin{align}
\label{eq:loss}
\mathcal{L}(E^{(K)}_{ij}) = \Big\langle \log \frac{\big\|\text{cycle $c$ end residual}\big\|}{\big\|\text{cycle $c$ start residual}\big\|} \Big\rangle_{c\,\in\,\{1,2,4,7,11\}} \, ,
\end{align}
where we have averaged over a set of five restart cycles, $c$ (solves that converge before restart 11 contribute only the cycles they require). For simplicity, the residuals at the beginning of each restart cycle come from an actual solve and are held fixed. Phrased differently, our loss function asks: ``how much does each 50-iteration GMRES cycle reduce the residual?''; this inclusion of GMRES knowledge within $\mathcal{L}$ opens the door to learning information about the eigenvalue spectrum of the preconditioned operator. Notice that the loss function is described in terms of the GNN output, $E^{(K)}_{ij}\rightarrow \mathbf{P}_\text{neural}$. To enable backpropagation, we need to calculate $\partial\mathcal{L}/\partial E^{(K)}_{ij}$, which is difficult because we have to differentiate through a sequence of GMRES iterations. GMRES builds, by the Arnoldi process, an orthonormal basis of the Krylov space spanned by ($\mathbf{r}$, $(\mathbf{A}\mathbf{P}^{-1})\mathbf{r}$, $(\mathbf{A}\mathbf{P}^{-1})^2 \mathbf{r}$, $\ldots$), 
and selects the combination of basis vectors that minimizes the residual by building a Hessenberg matrix and solving a least-squares problem. So we need to differentiate through a forward-pass sequence of triangular solve + matrix multiplication + orthogonalization + normalization + least-squares operations corresponding to the sequence of GMRES iterations.

We calculate these $\partial\mathcal{L}/\partial E^{(K)}_{ij}$ values by manual backward-pass differentiation of each restart cycle (manual rather than automatically through \texttt{pytorch} features): the derivative of the least-squares residual with respect to the Hessenberg matrix is propagated backwards through the normalization and Gram-Schmidt orthogonalization steps of the Arnoldi process, through $\mathbf{A}^{\dagger}$, and through the adjoint preconditioner solves $(\mathbf{P}_\text{neural}^{-1})^\dagger$. At each Arnoldi step, the sensitivity to the factorized entries of $\mathbf{P}_\text{neural}$ is an outer product of a forward and an adjoint vector, evaluated on the sparsity pattern of $\mathbf{A}$. Differentiating one GMRES restart cycle to produce $\partial\mathcal{L}/\partial E^{(K)}_{ij}$ costs roughly twice as much as running that restart cycle forward during a solve because it involves a forward-pass followed by a backward-pass. Finally, the resulting gradient is then back-propagated through the GNN with \texttt{pytorch}.

\begin{figure}
\includegraphics[trim={0 7 0 0},clip,width=\columnwidth]{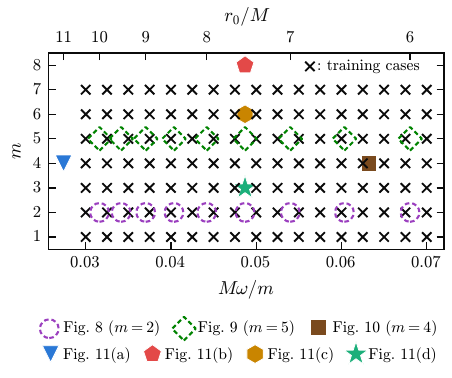}
\caption{Training and test cases for the black hole perturbation problem in the operator parameter space of $m$-mode and mode frequency $\omega$. The black crosses mark the 17 training cases for each of $m=1,\ldots,7$, spanning $M\omega/m=0.03$ to $0.07$ in steps of $0.0025$; a separate GNN is trained for each $m$. For circular orbits the mode frequency is $\omega=m\Omega$ with $M\Omega=(M/r_0)^{3/2}$, so the bottom axis maps to the orbital radius $r_0/M$ on the top axis, and the training set covers $5.9\lesssim r_0/M\lesssim 10.4$. Colored symbols mark the cases presented in later figures: the nine orbits $r_0/M=6, 6.5, \ldots, 10$ for $m=2$ (\cref{fig:summary_m2}) and $m=5$ (\cref{fig:summary_m5}), which fall between the training frequencies; the resolution study at $m=4$ and $r_0=6.3M$ (\cref{fig:resolution}); and the four extrapolation tests of \cref{fig:generalize}, which lie outside the training set in frequency ((a),~$r_0=11M$), in $m$ value ((b),~$m=8$), in resolution ((c),~2.1$\times$ unknowns), and in source vector ((d),~random~$\mathbf{b}$).}
\label{fig:train} 
\end{figure}

Although these gradients are more expensive to evaluate than those of the reference loss function candidates that were unsuccessful for our systems, training costs are manageable; no precomputed target factorizations are needed, because the loss is produced by GMRES itself. Because $m$ is a discrete parameter, we train separate GNNs for each $m$. For each $m$, the training set consists of 17 frequencies with $M\omega/m = (0.03, 0.0325, 0.035, \ldots, 0.07)$, which corresponds to $5.9 \lesssim r_0/M \lesssim 10.4$, see Fig.~\ref{fig:train}. For proof-of-concept, we perform each training at fixed resolution such that there are a total of approximately 190,000 unknown field values. A set of cases that are not part of the training set are used for validation. Each weight update uses a single training frequency: a GMRES solve with the current network supplies the restart-cycle residuals, after which the loss gradient is computed and applied with the Adam optimizer~\cite{1371975839702747014} (with learning rate $3\times10^{-4}$, linear warm-up over the first 8 updates, and gradient-norm clipping at 1). An epoch is one pass over the training frequencies in random order. After each epoch, the total number of GMRES iterations on the held-out cases is evaluated; an epoch that fails to reduce it by at least 0.5\% is reverted and the learning rate halved, and training stops after three such epochs (typically 6-22 epochs in total). The $m=2$ network was trained first, and then each other $m$ was initialized from the trained network of a neighboring $m$, which shortens training and, for $m=6$, was necessary to obtain any convergence at all when $r_0 = 6M$. Training on a laptop CPU costs approximately 4 hours per $m$, and inference on a laptop CPU requires approximately 2 seconds.

\section{Results}
\label{sec:results}

\subsection{Neural acceleration of black hole perturbations}

\subsubsection{Performance of neural preconditioner}

\begin{figure}
\includegraphics[trim={0 7 0 0},clip,width=\columnwidth]{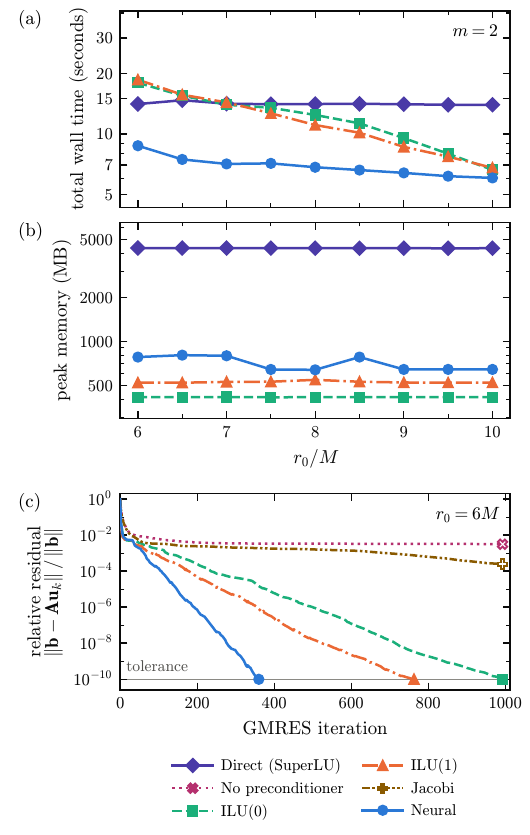}
\caption{Performance of the black hole perturbation $\mathbf{P}_\text{neural}$ compared to traditional methods for $m=2$ at the nine test orbits (purple circles in \cref{fig:train}), which are not in the training set. (a) Total wall time on a laptop CPU (including inference) and (b) peak memory are shown vs.~$r_0$; only converged methods are included. The neural preconditioner is the fastest method across $r_0$ and requires roughly 7$\times$ less memory than the direct solver, although slightly more than the ILU preconditioners. (c) Relative residual versus GMRES iteration for the highest-frequency (and most difficult) case, $r_0=6M$, is shown. The neural preconditioner converges (horizontal gray line) in less than half as many iterations as ILU(1) and a third as many as ILU(0), while Jacobi and no-preconditioner stall.}
\label{fig:summary_m2} 
\end{figure}

\begin{figure}
\includegraphics[trim={0 7 0 0},clip,width=\columnwidth]{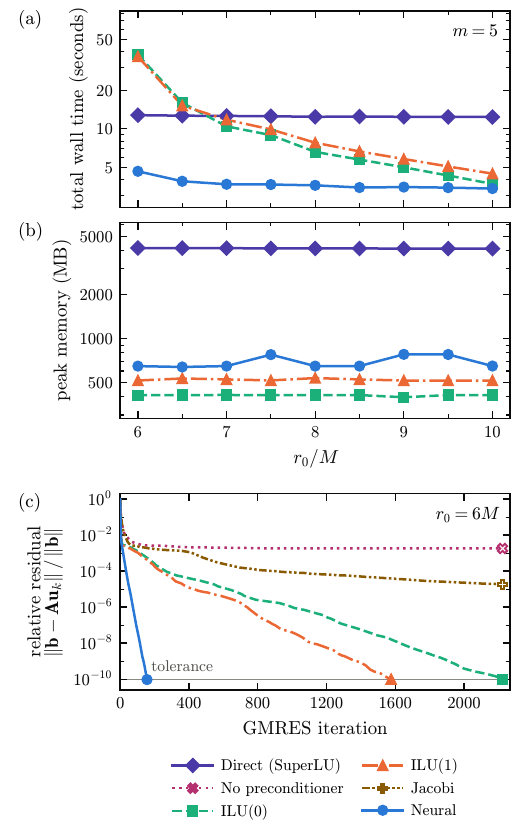}
\caption{Same as \cref{fig:summary_m2}, but for $m=5$ (green diamonds in \cref{fig:train}). The neural preconditioner is especially impressive for these cases: at $r_0=6M$ it reaches the
tolerance (horizontal gray line) with $10\times$ fewer iterations than ILU(1) and $15\times$ fewer than ILU(0), and it has the lowest wall time of all methods at every $r_0$, including roughly a $3\times$ speedup over the direct solver at
$r_0=6M$ (and even better than that for other~$r_0$).}
    \label{fig:summary_m5} 
\end{figure}

Figures~\ref{fig:summary_m2} and~\ref{fig:summary_m5} demonstrate that the neural preconditioner method is faster than all other methods we identified to solve this formulation~\cite{osburn2026schwarzschildperturbationslorenzgauge} of the spacetime metric perturbation PDEs. For lower frequency cases, the neural preconditioner produced a modest gain of approx. $1.5\times$ speed. In contrast, cases involving higher frequencies were accelerated considerably by the neural preconditioner. For these high-frequency cases, even defeating the direct matrix solver was not possible without the neural preconditioner, which enabled 2-15$\times$ fewer iterations than basic ILU and approximately 2-4$\times$ less wall time than the next best method (on a laptop CPU). We also observed that certain high frequency cases involving a near-zero pivot, such as $m=6$ at $r_0/M=6$, caused the traditional ILU methods to fail entirely. Impressively, and despite training a correction to ILU(0), the neural preconditioner naturally learns how to repair these near-zero pivots that otherwise would have forced usage of a direct solver. 

\begin{figure}
\includegraphics[trim={0 7 0 0},clip,width=\columnwidth]{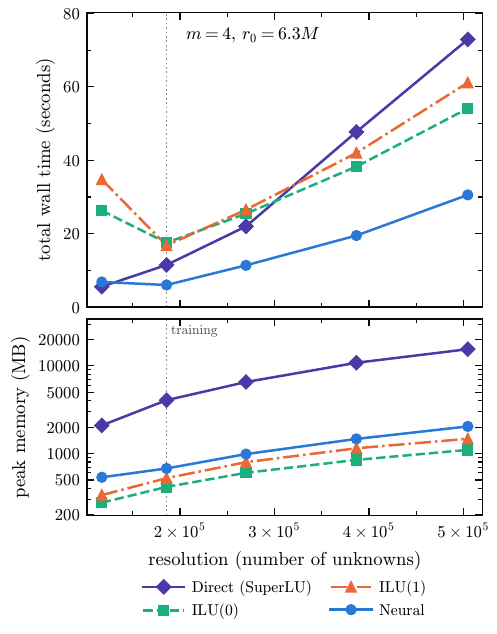}
\caption{Generalization of the neural preconditioner to resolutions outside the training is shown for $m=4$ and $r_0=6.3M$ (brown square in \cref{fig:train}). The network was trained only at the resolution marked by the vertical dotted line ($1.8\times10^5$ unknowns) and applied without retraining from $1.2\times10^5$ to $5\times10^5$ unknowns. The top shows total wall time vs. resolution and the bottom shows peak memory vs. resolution. The neural preconditioner remains faster than ILU(0) and ILU(1) at every resolution and faster than the direct solver at all but the lowest resolution. The memory required by the direct solver grows steeply with resolution, reaching nearly $20$\,GB at the highest resolution shown. The original direct solver implementation~\cite{osburn2026schwarzschildperturbationslorenzgauge} required even higher resolutions and consumed hundreds of GB on a large workstation to solve the problem. Although the time required also grows rapidly with resolution, the true weakness of the direct solver is unrealistically high memory requirements. }
\label{fig:resolution} 
\end{figure}

These Fig.~\ref{fig:summary_m2} and~\ref{fig:summary_m5} comparisons give the direct solver a fighting chance by considering only a modest resolution. Pursuing the highest resolutions reported in Ref.~\cite{osburn2026schwarzschildperturbationslorenzgauge} causes a blowup in the memory and time required by the direct solver, see \cref{fig:resolution}. Notice how at higher resolutions we tested cases where the direct solver requires nearly 20 GB of memory; furthermore, the original direct solver implementation~\cite{osburn2026schwarzschildperturbationslorenzgauge} required even higher resolutions and consumed hundreds of GB on a large workstation to solve the problem. Although the time required also grows rapidly with resolution, the true weakness of the direct solver is unrealistically high memory requirements. By outperforming both the direct and traditional iterative methods across parameter space, our neural preconditioner enables elimination of the direct solver (and its associated weaknesses) entirely.

\subsubsection{Testing generalizability beyond training domain}
\label{sec:generalize}

One desirable feature we targeted for the neural preconditioner is generalizability beyond the training domain. If the training time cost were comparable to the time saved during implementation, it would defeat the purpose; therefore, it is important to investigate how plausible it is to learn from a smaller set of training cases (and minimize training time) without harming performance of the neural preconditioner. First consider whether the neural preconditioner is able to generalize across a broad range of resolutions. We stress-tested this feature to the highest extent by training at a fixed resolution and then deploying the neural preconditioner to solve problems across many resolutions. Figure~\ref{fig:resolution} demonstrates that the neural preconditioner successfully generalizes to significantly higher resolutions than it was trained on originally. Arguably this choice is much stricter than necessary because we could have included multiple resolutions in the training set, which makes our successful resolution generalization that much more impressive.

\begin{figure*}[t]
\includegraphics[trim={5 7 3 0},clip,width=\textwidth]{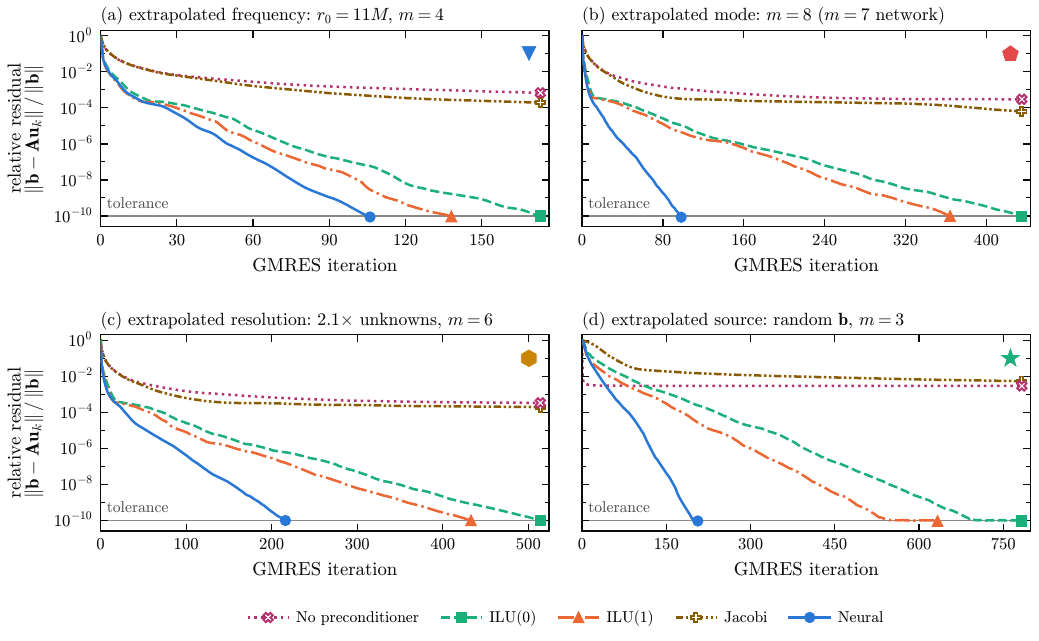}
\caption{Extrapolations of the neural preconditioner beyond the training domain are shown for the black hole perturbation problem. The upper-right symbols correspond to the legend of \cref{fig:train}. The following extrapolations are depicted: (a) extrapolation to a lower frequency, $r_0=11M$, for $m=4$; (b) extrapolation to $m=8$ using the network trained for $m=7$; (c) extrapolation to $2.1\times$ the number of unknowns, for $m=6$; and (d) extrapolation to a source vector $\mathbf{b}$ populated with random numbers (scaled relative to the diagonal) rather than the pattern used in training, for $m=3$ (the randomness causes a near-stall approaching the tolerance). Panels (b)-(d) use $r_0=7.5M$. In every case the neural preconditioner reaches the $10^{-10}$ tolerance in the fewest iterations, between approx. $1.3\times$ fewer for (a) and $4\times$ fewer for (b) compared to the next best method, ILU(1). The success in (d) indicates that the network has learned the homogeneous PDE operator itself rather than learning restricted pairs with each operator matched to a specific source.}
\label{fig:generalize} 
\end{figure*}

Other dimensions of generalizability include extrapolating $\omega$, extrapolating $m$, and extrapolating the source vector to categories outside what was used during training. Figure~\ref{fig:generalize} tests these new dimensions of generalizability and also includes an additional example of extrapolated resolution. We trained the preconditioner across a finite range of $\omega$, and \cref{fig:generalize}(a) extrapolates to a lower frequency outside the training domain. Similarly, we trained the preconditioner across a finite range of $m$, and \cref{fig:generalize}(b) extrapolates to a higher $m$ outside the training domain. Figure~\ref{fig:generalize}(c) extrapolates to 2.1$\times$ the number of unknowns for a pair of $m$ and $r_0$ independent of \cref{fig:resolution}, and Fig.~\ref{fig:generalize}(d) populates the source vector with random numbers to contrast the circular motion source vectors used for training. Because training exclusively used circular motion source vectors, there may have been a risk of memorizing the restricted class of provided sources, which would defeat the important goal of extending to generic orbital motion; our strong performance with a random source vector implies that our training and loss strategy produces a neural preconditioner that has learned the homogeneous PDE operator itself rather than learning restricted pairs with each operator matched to a specific source. Note that the neural preconditioner outperforms traditional iterative methods for all extrapolations depicted in \cref{fig:generalize}.

\subsubsection{Analysis of the learned preconditioner}

\begin{figure}
\includegraphics[trim={3.5 5 0 0},clip,width=\columnwidth]{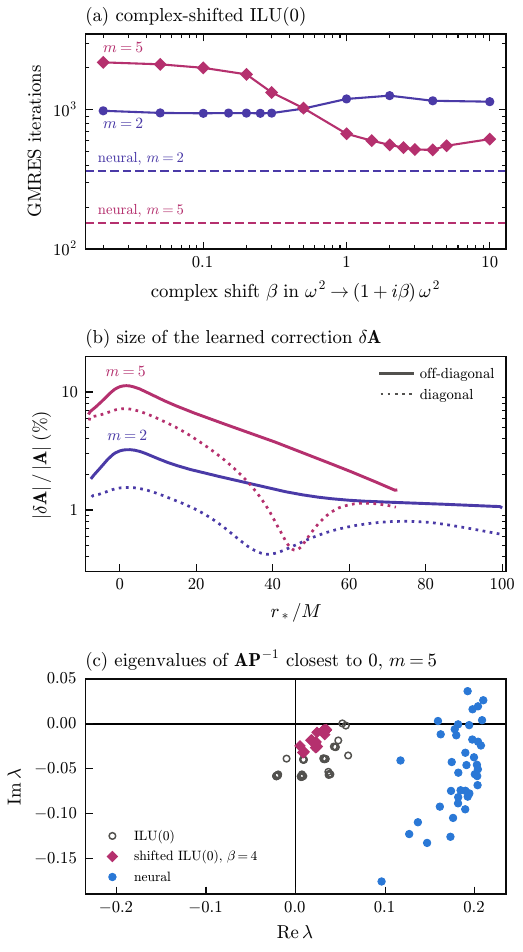}
\caption{Analysis of the learned preconditioner for the black-hole perturbation problem at $m=2$ (\cref{fig:summary_m2}) and $m=5$ (\cref{fig:summary_m5}) at $r_0=6M$. Panel (a) compares the neural preconditioner to a complex-shifted ILU(0). The complex shift reduces the GMRES iteration count substantially in the more oscillatory $m=5$ case at an optimal $\beta \approx 4$, but does not in the $m=2$ case. The neural preconditioner outperforms it in both cases. Panel (b) shows the learned correction $\delta\mathbf{A}$ relative to $\mathbf{A}$. At each point in $(r_*,\theta)$, the norm over the entries corresponding to the ten tensor components is taken, and averaged over $\theta$. Diagonal and off-diagonal entries are shown separately, demonstrating that the neural preconditioner applies its correction primarily through spatially dependent off-diagonal couplings. Panel (c) shows the 40 eigenvalues of the preconditioned operator closest to the origin, confirming that the complex-shifted ILU(0) moves the eigenvalues into the right half-plane but leaves them close to zero, while the neural preconditioner moves them further away from the origin and into the right half-plane.}
\label{fig:mechanism} 
\end{figure}

We can extract the learned $\delta\mathbf{A}$ from the trained network by \cref{eq:Pneural} to analyze its structure and understand how it improves the preconditioning.
First, it is clear from its construction, \cref{eq:Pneural}, that the neural preconditioner is not merely a more accurate factorization of $\mathbf{A}$ within the allowed sparsity pattern, but that it is equivalent to ILU(0) applied to a corrected operator $\mathbf{A}+\delta\mathbf{A}$.

The natural hypothesis is that $\delta\mathbf{A}$ could be a complex shift, a classic remedy for Helmholtz-type problems discussed in \cref{sec:linear}. We test this directly in \cref{fig:mechanism}(a) by comparing the neural preconditioner to a complex-shifted ILU(0). However, we find that the neural preconditioner outperforms the simple complex-shifted ILU(0) considerably. To understand this, we inspect the learned $\delta\mathbf{A}$ in \cref{fig:mechanism}(b). We find that $\delta\mathbf{A}$ is not a simple complex shift of the diagonal, but rather a spatially dependent complex correction that is applied primarily through off-diagonal local couplings within the sparsity pattern of the original operator. This structure is learned by training on the GMRES residuals, which are sensitive to the eigenvalue spectrum of the preconditioned operator. This eigenvalue spectrum is shown in \cref{fig:mechanism}(c), where the neural preconditioner moves the eigenvalues further away from the origin and into the right half-plane, which is a more favorable configuration for GMRES convergence. The complex-shifted ILU(0) also moves the eigenvalues into the right half-plane, but leaves them close to zero. Therefore, the neural preconditioner is able to learn a correction that improves the convergence of GMRES beyond what a simple complex shift can achieve.

The neural preconditioner does not, however, learn to emulate a global preconditioner like multigrid, which is able to solve for large-scale modes in the solution, because its construction is local and only exchanges information between nearby grid points. Therefore, it does not eliminate the growth of the iteration count with resolution, as demonstrated in \cref{sec:generalize}.

\subsection{Neural acceleration of elastodynamics}
\label{sec:elasto_accel}

Our cross-disciplinary elastodynamics implementation largely follows the same neural preconditioner logic as our primary gravitational application; here we summarize any differences between the two models and highlight the strong performance of the neural preconditioner in this other area. One difference of the elastodynamics GNN is the content of the initial node features. Arguably this is the area where different applications require the most customization, and we choose the initial node feature vector to be $\vec{v}^{\,(0)}_i = (z_0, \kappa_0, r_i, z_i, f^\text{comp}_i, f^\text{bound}_{r_{i}}, f^\text{bound}_{z_i})$, where $f^\text{comp}_i$ is a two-state flag indicating the vector component ($r$ or $z$) of $\vec{u}$ and $f^\text{bound}_{r_{i}/z_i}$ are three-state flags indicating whether $r_{i}$ or $z_i$ is at a boundary and, if so, which boundary. Another difference with our elasticity application is we train $\mathbf{P}_\text{neural}$ as a correction to ILU(1) rather than ILU(0); for elasticity we observed a performance gain by correcting ILU(1), which differs from what we observed for the gravitational case, where no significant gains were produced by correcting ILU(1) compared to ILU(0).

\begin{figure}
\includegraphics[trim={2 7 5 2},clip,width=\columnwidth]{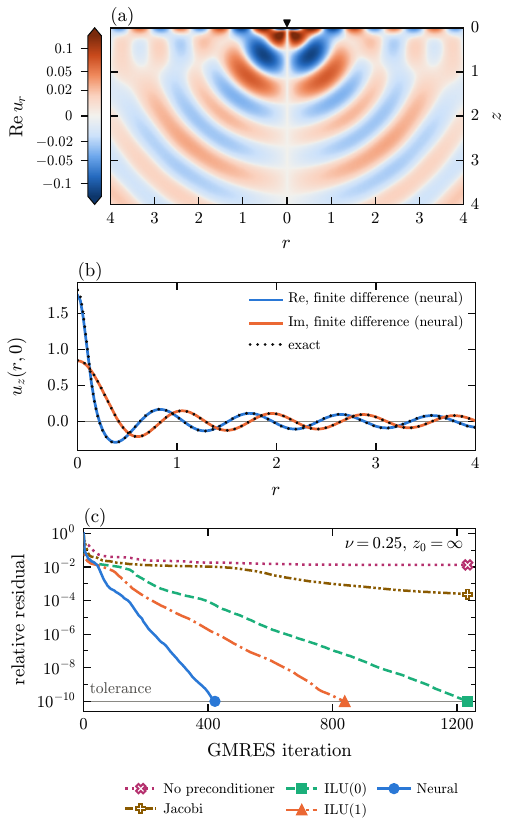}
\caption{Elastic waves sourced at the origin in a half-space are shown for the homogeneous case ($z_0=\infty$) with Poisson ratio $\nu=0.25$. The following features are depicted: (a) the real part of the radial displacement $u_r$ in the $r$-$z$ plane (the triangle is the center of the surface pressure); (b) vertical surface displacement $u_z(r,0)$ comparison between the neural preconditioner numerical solution and the exact solution (which is available for this simpler case); (c) relative residual versus GMRES iteration, where the neural preconditioner converges in approximately half as many iterations as ILU(1). All quantities are in the dimensionless units of \cref{sec:elasto}.}
\label{fig:lamb} 
\end{figure}

\begin{figure}
\includegraphics[trim={2 7 5 2},clip,width=\columnwidth]{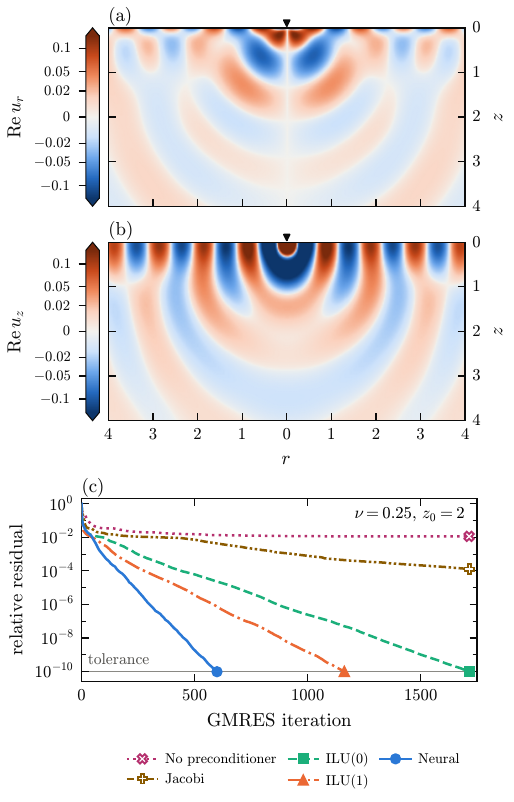}
\caption{Similar to \cref{fig:lamb}, but for a non-homogeneous half-space with $z_0=2$ (so $\mu$ and $\lambda$ double their surface values at depth $z=2$) and with $u_z$ shown across the whole domain because no exact solution is available. Because the wave speeds increase with depth, the wavelengths grow with $z$, stretching the wave pattern at depth relative to \cref{fig:lamb}. As in \cref{fig:lamb}, the neural preconditioner converges in approximately half as many iterations as ILU(1).}
\label{fig:elasticity} 
\end{figure}

Figures~\ref{fig:lamb} and~\ref{fig:elasticity} demonstrate the strong performance of the neural preconditioner compared to traditional iterative methods for elastic waves. Figure~\ref{fig:lamb} considers a homogeneous medium by having $z_0=\infty$, and for which there is an exact solution to compare with. Figure~\ref{fig:elasticity} considers a non-homogeneous medium by having $z_0 = 2$. The same GNN was used successfully across both cases, which demonstrates reliable interpolation of $\mathbf{P}_\text{neural}$ within the parameter space ($z_0$, $\kappa_0$). Note that this simpler scenario involves a much smaller number of unknown field values, which is the regime where direct matrix solvers thrive; for this reason the direct solver was slightly faster in \cref{fig:lamb,fig:elasticity}, although the neural preconditioner outperformed all other iterative methods. We predict the additional degrees of freedom associated with 3D elasticity problems will be able to appreciate the benefits of a neural preconditioner even more deeply than this simpler 2D axisymmetric example.

\section{Conclusions and future work}
\label{sec:conclusions}

Machine learning and high accuracy need not be in conflict. We have shown how neural networks can be trained to substantially accelerate the solution of demanding elliptic PDEs in gravitational physics and beyond, while preserving final accuracy according to a traditional Krylov residual rather than exposing the final solution to the noise-floor of the network. The key design choice is where the network is placed: rather than predicting the solution directly, our GNN produces the preconditioner for a traditional iterative algorithm, which measures the true residual at every iteration and continues until the prescribed tolerance is satisfied, see \cref{fig:idea}. Any small imperfections that may linger in our learned preconditioner therefore never cost accuracy.

For the ten coupled Lorenz gauge PDEs governing the $m$-modes of the first-order spacetime metric perturbation~\cite{ellipticgsf_neural}, the neural preconditioner converges to relative residuals of $10^{-10}$ with 2-15$\times$ fewer GMRES iterations than ILU and 2-4$\times$ less wall time than the next best method for the most challenging high-frequency cases (\cref{fig:summary_m2,fig:summary_m5}). It also eliminates the need for a memory-intensive direct solver, whose requirements grow steeply with resolution (\cref{fig:resolution}), and it repairs the near-zero pivots that can cause traditional ILU to fail entirely.

Two features of our approach proved essential. First, the loss function: the Frobenius-norm losses of \ccites{hausner2024,hausner2024GMRES} produced preconditioners that were worse than ILU(0) for our indefinite Helmholtz-type systems, whereas our novel loss directly measures the residual reduction of each GMRES restart cycle, with gradients obtained by differentiating through the Arnoldi process. Training against the downstream solver objective, rather than a proxy for factorization accuracy, allows the network to learn about the eigenvalue spectrum of the preconditioned operator. Indeed, our analysis of the trained network (\cref{fig:mechanism}) shows that it has learned to factorize a modified operator, $\mathbf{A}+\delta\mathbf{A}$, whose spatially dependent and predominantly off-diagonal complex correction moves the eigenvalues of the preconditioned operator away from the origin and into the right half-plane, outperforming the traditional complex-shift remedy. Second, the preconditioner generalizes beyond its training domain in frequency, $m$, resolution, and source pattern (\cref{fig:resolution,fig:generalize}). The random-source test is particularly significant because it indicates that the network has learned the homogeneous PDE operator itself rather than specific operator-source pairs; this feature enables certain valuable applications described below.

The same strategy transferred to elastodynamics with only minor customization (different initial node features and an ILU(1) baseline), where a single network required approximately half as many GMRES iterations as ILU(1) for both the homogeneous Lamb's problem and a non-homogeneous half-space (\cref{fig:lamb,fig:elasticity}). Because the GNN operates directly on the sparse matrix with only modest physical context, new applications are enabled by retraining rather than manual redesign of the preconditioner. We expect the advantages of the neural preconditioner to become even more prominent for larger 3D elasticity problems where direct methods scale poorly. Elliptic elasticity problems also arise within gravitational wave science itself; for example, Brownian thermal noise calculations for detector mirrors are already performed with SpECTRE~\cite{Vu:2021mhw}.

Our results also clarify the region of highest improvement for the present construction. Each element of $\mathbf{P}_\text{neural}$ depends only on matrix elements within $K=4$ grid points, so the neural preconditioner cannot emulate the global coarse-scale corrections of multigrid; thus, the iteration count still grows with resolution. At low frequencies, where few iterations are required, inference becomes a non-negligible fraction of the total wall time and the advantage over ILU narrows; GPU inference should reduce this overhead. Finally, the GNN requires an explicitly assembled sparse matrix, whereas many methods only require the matrix-vector product. Both the locality and explicit matrix assembly issues are naturally addressed by the multigrid setting discussed below.

The most important application of neural preconditioning in this context is the production of first-order metric perturbation data for generic orbits around Kerr black holes, which is the essential input for constructing sources for second-order self-force calculations. Second-order self-force calculations are required for accurate LISA parameter estimation~\cite{Hinderer_2008,Burke_2024}, and the second-order source is quadratic in the first-order metric perturbation and its derivatives. Constructing it therefore requires accurate first-order $m$-modes throughout the 2D domain for every mode that participates in the mode coupling, which for generic Kerr orbits amounts to an enormous number of elliptic solves. A generic bound orbit is specified by the black hole spin $a$, the semi-latus rectum $p$, the eccentricity $e$, and the inclination parameter $x$, and its source decomposes into harmonics labeled by $(m,k,n)$ with frequencies $\omega_{mkn}=m\Omega_\varphi+k\Omega_\theta+n\Omega_r$. The source parameter space is thus seven-dimensional: $(a,p,e,x,m,k,n)$. In contrast, the homogeneous operator of the frequency-domain $m$-mode field equations depends only on $(a,\omega,m)$~\cite{nishimura2026}. Because our neural preconditioner represents the homogeneous operator alone, this collapse from a 7D source parameter space to a 3D operator parameter space enables efficient training of a compact family of neural preconditioners: one network per $m$, trained over the continuous $(a,\omega)$ domain with $a$ added to the initial node features of \cref{eq:initNode}. Thus, such a neural preconditioner can serve every orbit and every harmonic across the 7D source parameter space. The training cost of each network is then amortized over all $(p,e,x,k,n)$ combinations that map into its domain rather than repeated for each orbit. The same family also applies to the second-order solves themselves, because the second-order field equations involve the same linear operator acting on the second-order metric perturbation, with frequencies and azimuthal numbers given by combinations of those of the coupled first-order modes. Given that training on a single laptop CPU currently requires approximately 4 hours per $m$ across a 1D frequency domain, we anticipate that covering the 2D $(a,\omega)$ domain will remain a modest one-time cost compared with the volume of solves it accelerates.

Our recent spectral and discontinuous Galerkin implementations of the frequency-domain $m$-mode method~\cite{Vu_2026,nishimura2026} provide an immediate route to deploying neural preconditioners at scale. There, the elliptic solves use multigrid with a Schwarz smoother~\cite{Vu_2022}, and the Schwarz smoother step, which assembles a sparse matrix and computes its full LU factorization for every subdomain on every multigrid level, is the most costly step of the solve. This step is an excellent candidate for GNN preconditioning and the associated speed enhancement. Smoothers need only be approximate, so an exact LU factorization is more than is necessary; the subdomain matrices are small, structurally similar, and repeated across elements, multigrid levels, frequencies, and modes, so each trained network would be amortized over an exceptionally large number of inferences; and because the GNN operates on the sparsity graph rather than on a fixed stencil, it can accommodate the dense block structure of spectral elements. Moreover, embedding a local neural preconditioner as the smoother within multigrid directly addresses its main limitation: the coarse multigrid levels supply the global corrections that the GNN cannot, which should curb the growth of the iteration count with resolution. Because explicit assembly of the subdomain matrices is itself a dominant expense, matrix-free variants that act directly on residuals, such as neural operators~\cite{Li_2021}, which are a natural match for spectral and discontinuous Galerkin representations, are a compelling future direction.

More broadly, the indefinite and coupled systems considered here are representative of the many fields identified in \cref{sec:intro} where effective preconditioners remain elusive, including seismic full-waveform inversion~\cite{Virieux_2009}, nanophotonic inverse design~\cite{Molesky_2018}, implicit fusion plasma simulations~\cite{Jardin_2012}, and numerical relativity initial data~\cite{Vu_2022}. Neural preconditioning is also complementary to other ML enhancements of iterative solvers~\cite{meng2026overviewmachinelearningenhancediterative}, such as learned initial guesses~\cite{eshaghi2026nowsneuraloperatorwarm}, neural operators embedded within Krylov methods~\cite{pmlr-v235-rudikov24a}, and GNN-based learned linear algebra~\cite{trifonov2025learninglinearalgebragraph}. In each case the central lesson of this work applies: by placing fast-but-noisy ML as an accelerator where imperfection is tolerated and leaving accuracy validation to a traditional algorithm, fields with stringent accuracy requirements, such as gravitational wave source modeling for LISA, can benefit from ML without sacrificing the high precision they need.

\begin{acknowledgments}
TO gratefully acknowledges support from the National Science Foundation under Grant No.\ PHY-2309020 at SUNY Geneseo and from the Fulbright U.S. Scholar Program, which is sponsored by the U.S. Department of State and the Fulbright Commission in Ireland; the contents are solely the responsibility of the authors and do not necessarily represent the official views of the Fulbright Program, the U.S. Government, or the Fulbright Commission in Ireland.
NLV gratefully acknowledges support from the Swiss National Science Foundation (SNSF) Ambizione Grant No.~PZ00-2\_232961. For the purpose of open access, a CC BY public copyright license is applied to any author accepted manuscript (AAM) version arising from this submission.

\textit{AI usage:}
Claude Opus by Anthropic (multiple versions, most recently 5.5) was used for migrating the original Mathematica code~\cite{osburn2026schwarzschildperturbationslorenzgauge} to Python~\cite{ellipticgsf_neural}, debugging, performance optimization, unit testing, plot assembly, expanding our literature search, and writing/structure suggestions. All intellectual contributions, including the high-level ideas, development, and writing, are the original work of the authors.
All AI outputs were independently verified by the authors, who take full responsibility for the accuracy and content of this paper.
\end{acknowledgments}

\appendix

\section{Schur complement and boundary strips}
\label{sec:schur}

\begin{figure}
\includegraphics[width=0.8\columnwidth]{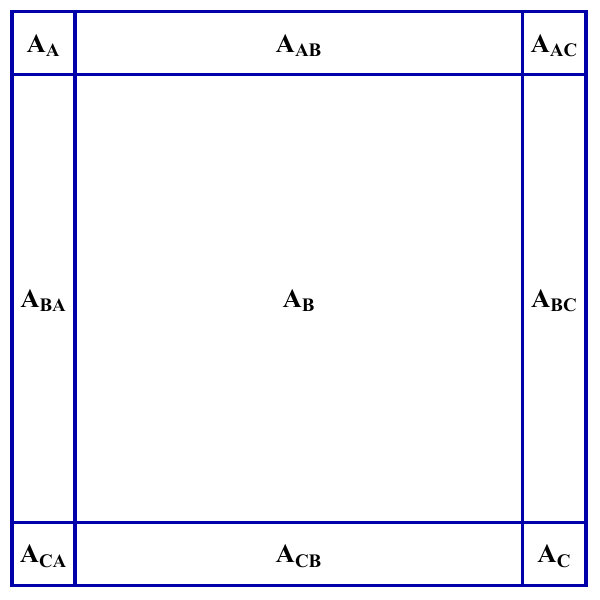}
\caption{Block partition of $\mathbf{A}$ (not drawn to scale). The top horizontal strip, $(\mathbf{A}_{A},\mathbf{A}_{AB},\mathbf{A}_{AC})$, involves extended radial coupling from the near-horizon boundary condition, and similar for the bottom horizontal strip, $(\mathbf{A}_{CA},\mathbf{A}_{CB},\mathbf{A}_{C})$, which to a lesser degree involves extended radial coupling from the outer radial boundary condition (see \cref{fig:matrix}). Algebraically eliminating the radial boundary unknowns and constructing a slightly smaller linear system involving the Schur complement, $\mathbf{A}_S$ (with same dimensions as $\mathbf{A}_B$), enables much better convergence properties for iterative methods.}
\label{fig:schur} 
\end{figure}

The original sparse matrices, $\mathbf{A}$, have very poor convergence properties with all iterative methods we attempted for both black hole perturbations and elasticity. We determined the cause to be the higher derivatives that appear in certain boundary conditions, see Fig.~\ref{fig:matrix}. Specifically, overcoming a near-horizon issue required imposing a radiative boundary condition involving the fifth $r_*$ derivative near the horizon, which with a one-sided finite difference stencil couples the seven radii closest to the horizon; similarly, the boundary condition at large radius involves the third $r_*$ derivative, and couples the five nearest radii, see Fig.~\ref{fig:schur}. The relatively small number of radii directly coupled to the boundaries are confined within thin strips of $\mathbf{A}$ that involve small subsets of unknown field values $\mathbf{u}_A$ and $\mathbf{u}_C$, while the bulk of the unknowns are contained within $\mathbf{u}_B$:
\begin{align}
&\left(\begin{array}{ccc} \mathbf{A}_{A} & \mathbf{A}_{AB} & \mathbf{A}_{AC} \\ \mathbf{A}_{BA} & \mathbf{A}_{B} & \mathbf{A}_{BC} \\ \mathbf{A}_{CA} & \mathbf{A}_{CB} & \mathbf{A}_{C}
\end{array}\right) \left(\begin{array}{c} \mathbf{u}_{A} \\ \mathbf{u}_{B} \\ \mathbf{u}_{C}
\end{array}\right) = \left(\begin{array}{c} \mathbf{b}_{A} \\ \mathbf{b}_{B} \\ \mathbf{b}_{C}
\end{array}\right) \, .
\end{align}
It is straightforward to rearrange the block matrix to an equivalent form where the boundary strips are adjacent 
\begin{align}
&\left(\begin{array}{ccc} \mathbf{A}_{A} & \mathbf{A}_{AC}  & \mathbf{A}_{AB} \\ \mathbf{A}_{CA} & \mathbf{A}_{C} & \mathbf{A}_{CB} \\ \mathbf{A}_{BA} & \mathbf{A}_{BC} & \mathbf{A}_{B}
\end{array}\right) \left(\begin{array}{c} \mathbf{u}_{A} \\ \mathbf{u}_{C} \\ \mathbf{u}_{B}
\end{array}\right) = \left(\begin{array}{c} \mathbf{b}_{A} \\ \mathbf{b}_{C} \\ \mathbf{b}_{B}
\end{array}\right) \, .
\end{align}
It is convenient to combine $\mathbf{u}_A$ and $\mathbf{u}_C$ as one set of boundary strip unknowns, $\mathbf{u}_0$
\begin{align}
& \mathbf{u}_0 \equiv \left(\begin{array}{c} \mathbf{u}_{A} \\ \mathbf{u}_{C} \end{array} \right) \, ,
\\
& \mathbf{b}_0 \equiv \left(\begin{array}{c} \mathbf{b}_{A} \\ \mathbf{b}_{C} \end{array}\right) \, ,
\\
& \mathbf{A}_0 \equiv \left(\begin{array}{cc} \mathbf{A}_{A} & \mathbf{A}_{AC} \\ \mathbf{A}_{CA} & \mathbf{A}_{C}
\end{array}\right) \, ,
\\
& \mathbf{A}_{0B} \equiv \left(\begin{array}{c} \mathbf{A}_{AB} \\ \mathbf{A}_{CB} 
\end{array}\right) \, ,
\\
& \mathbf{A}_{B0} \equiv \left(\begin{array}{cc} \mathbf{A}_{BA} & \mathbf{A}_{BC} 
\end{array}\right) \, ,
\end{align}
and to rewrite the block representation of $\mathbf{A}$ as a $2\times 2$ block matrix
\begin{align}
&\left(\begin{array}{cc} \mathbf{A}_{0} & \mathbf{A}_{0B}\\ \mathbf{A}_{B0} & \mathbf{A}_{B} 
\end{array}\right) \left(\begin{array}{c} \mathbf{u}_{0} \\ \mathbf{u}_{B}
\end{array}\right) = \left(\begin{array}{c} \mathbf{b}_{0} \\ \mathbf{b}_{B}
\end{array}\right) \, .
\end{align}
It is straightforward to express $\mathbf{u}_{0}$ in terms of $\mathbf{u}_{B}$ and use substitution to derive a matrix equation governing $\mathbf{u}_{B}$ alone
\begin{align}
\mathbf{u}_{0} = \mathbf{A}_{0}^{-1} & \left( \mathbf{b}_{0} - \mathbf{A}_{0B} \mathbf{u}_{B} \right) \, , \label{eq:exact}
\\
\left(\mathbf{A}_B - \mathbf{A}_{B0}\mathbf{A}_{0}^{-1} \mathbf{A}_{0B} \right) & \mathbf{u}_{B} = \mathbf{b}_B - \mathbf{A}_{B0}\mathbf{A}_{0}^{-1}\mathbf{b}_{0}  \, .
\end{align}
The new sparse matrix multiplying $\mathbf{u}_B$ is the Schur complement, $\mathbf{A}_S$
\begin{align}
\mathbf{A}_S & \equiv \mathbf{A}_B - \mathbf{A}_{B0}\mathbf{A}_{0}^{-1} \mathbf{A}_{0B} \, ,
\\
\mathbf{b}_S & \equiv \mathbf{b}_B - \mathbf{A}_{B0}\mathbf{A}_{0}^{-1}\mathbf{b}_{0} \, ,
\end{align}
which produces the linear system governing $\mathbf{u}_B$
\begin{align}
& \mathbf{A}_S \mathbf{u}_B = \mathbf{b}_S \, . \label{eq:schur}
\end{align}
Because $\mathbf{u}_B$ is many times larger than $\mathbf{u}_0$, almost all of the computational cost is devoted to calculating $\mathbf{u}_B$. Thankfully we find that, after eliminating the radial boundary strips, $\mathbf{A}_S$ has much better convergence properties than $\mathbf{A}$ when solving via iterative methods; for this reason, the linear systems we mention in the main body of the text that are solved via iterative methods are always referring to Eq.~\eqref{eq:schur} (despite the misleading usage of $\mathbf{A}$, $\mathbf{u}$, and $\mathbf{b}$). After solving for $\mathbf{u}_B$ iteratively, it is a negligible cost to calculate $\mathbf{u}_0$ exactly via Eq.~\eqref{eq:exact}.

\bibliography{main}

\end{document}